\documentclass[a4paper,fleqn,usenatbib, useAMS]{mnras}
\usepackage{graphicx}
\usepackage{multicol}
\usepackage{enumerate}
\usepackage{amsmath}
\usepackage{pdflscape}
\usepackage{placeins}
\usepackage{amssymb}
\usepackage{longtable}
\usepackage[flushleft]{threeparttable}
\usepackage{journal_names}
\def\farcs{\hbox{$.\!\!^{\prime\prime}$}}
\def\kms{km s$^{-1}$}
\title[Optical spectroscopy of TDE host galaxies] {Black hole masses of tidal disruption event host galaxies II}
\author[T. Wevers et al.]{Thomas Wevers$^{1}$\thanks{Email: tw@ast.cam.ac.uk}, Nicholas C. Stone$^{2}$, Sjoert van Velzen$^{3,4}$, Peter G. Jonker$^{5,6}$, \newauthor Tiara Hung$^{4}$, Katie Auchettl$^{7,8,9}$, Suvi Gezari$^{4}$, Francesca Onori$^{5}$, Daniel \newauthor Mata S\'{a}nchez$^{10}$, Zuzanna Kostrzewa-Rutkowska$^{5,6}$ and Jorge Casares$^{11,12}$\\\\
$^{1}$Institute of Astronomy, Madingley Road, Cambridge CB3 0HA, United Kingdom\\
$^{2}$Columbia Astrophysics Laboratory, Columbia University, New York, NY 10027\\
$^{3}$Center for Cosmology and Particle Physics, New York University, New York, NY 10003\\
$^{4}$Department of Astronomy, University of Maryland, College Park, MD 20742\\
$^{5}$SRON, Netherlands Institute for Space Research, Sorbonnelaan 2, 3584CA Utrecht, The Netherlands\\
$^{6}$Department of Astrophysics/IMAPP, Radboud University, P.O. Box 9010, 6500GL Nijmegen, The Netherlands\\
$^{7}$Center for Cosmology and Astro-Particle Physics, The Ohio State University, 191 West Woodruff Avenue, Columbus, OH 43210, USA\\
$^{8}$Department of Physics, The Ohio State University, 191 W. Woodruff Avenue, Columbus, OH 43210, USA\\
$^{9}$Dark Cosmology Centre, Niels Bohr Institute, University of Copenhagen, Blegdamsvej 17, 2100 Copenhagen, Denmark\\
$^{10}$Jodrell Bank Centre for Astrophysics, School of Physics and Astronomy, The University of Manchester, Manchester M13 9PL, UK\\
$^{11}$Instituto de Astrof\'{i}sica de Canarias, E-38205 La Laguna, Tenerife, Spain \\
$^{12}$Departamento de Astrof\'{i}sica, Universidad de La Laguna, E-38206 La Laguna, Tenerife, Spain
}
\begin{document}
\date{\today}
\pagerange{\pageref{firstpage}--\pageref{lastpage}} \pubyear{2019}
\maketitle
\label{firstpage}
\begin{abstract}
We present new medium resolution, optical long-slit spectra of a sample of 6 UV/optical and 17 X-ray selected tidal disruption event candidate host galaxies. We measure emission line ratios from the optical spectra, finding that the large majority of hosts are quiescent galaxies, while those displaying emission lines are generally consistent with star-formation dominated environments; only 3 sources show clear evidence of nuclear activity. We measure bulge velocity dispersions using absorption lines and infer host black hole (BH) masses using the M\,--\,$\sigma$ relation. While the optical and X-ray host BH masses are statistically consistent with coming from the same parent population, the optical host $M_{\rm BH}$ distribution has a visible peak near $M_{\rm BH} \sim 10^6 M_\odot$, whereas the X-ray host distribution appears flat in $M_{\rm BH}$. We find a subset of X-ray selected candidates that are hosted in galaxies significantly less luminous (M$_{\rm g}$\,$\sim$\,--16) and less massive (stellar mass\,$\sim$\,10$^{8.5-9}$\,M$_{\odot}$) than those of optical events. Using statistical tests we find suggestive evidence that, in terms of black hole mass, stellar mass and absolute magnitude, the hard X-ray hosts differ from the UV/optical and soft X-ray samples. Similar to individual studies, we find that the size of the emission region for the soft X-ray sample is much smaller than the optical emission region, consistent with a compact accretion disk. We find that the typical Eddington ratio of the soft X-ray emission is $\sim$\,0.01, as opposed to the optical events which have L$_{\rm BB}$\,$\sim$\,L$_{\rm Edd}$. The latter seems artificial if the radiation is produced by self-intersection shocks, and instead suggests a connection to the SMBH. 
\end{abstract}


%%%%%%%%%%%%%%%%%%%%%%%%%%%%%%%%%%%%%%%%%%%%%%%%%%%%%%%%%%%%%%%%%%%%%%%%%%%%%%%%%%%%%%%%%%%%%%%%%%%%%%%%%%%%%%%%%%%%%%%%%%%%%%%%%%%%%%%%%%%%%%%%%%%%%%%%%%%%%%%%%%%%%%%%%%%%%%%%%%%%%%%%%%%%%%%%%%%%%%%%%%%%%%%%%%%%%%%%%%
\section{Introduction}
\label{sec:introduction}
The study of tidal disruption events (TDEs), where stars are torn apart by the immense tidal forces near supermassive black holes (SMBHs) in the centers of galaxies, has emerged in recent years as a new tool to study both dormant and active SMBHs. Basic theoretical predictions \citep{Hills1975, Rees1988, Evans1989, Phinney1989} were established two decades before the first observational claims of such events were made \citep{Grupe1995, Grupe1999}. Observations across the electromagnetic spectrum have since led to candidate TDE detections at almost every wavelength, including hard X-rays \citep{Bloom2011, Cenko2012, Walter2016}, soft X-rays (e.g. \citealt{Komossa1999, Greiner2000}), UV/optical (e.g. \citealt{Gezari2008,vanvelzen2011,Holoien2014,Arcavi2014}), infrared (IR; \citealt{vanvelzen2016b, Jiang2016, Matilla2018}), and radio waves \citep{vanvelzen2016a, Alexander2016}. So far the majority of well-established TDEs have been identified in UV/optical surveys, and in addition 3 jetted TDEs have been classified in hard X-rays \citep{Bloom2011, Cenko2012, Brown2015}. The soft X-ray TDE candidates have remained somewhat more ambiguous because of the intrinsic X-ray variability observed in active galactic nuclei (AGN), and an interpretation of these events as extreme AGN variability has not been completely ruled out (see e.g. \citealt{Auchettl2018}, who show that $\sim$\,1\,\% of AGN flares could resemble TDE emission). 
Another concern is the lack of temporal coverage and/or pre-flare X-ray limits that could rule out AGN activity. This leaves open the possibility that these flares are simply the extreme tail of normal AGN variability.

Although it is important to consider alternate explanations for large outbursts occurring in the centers of galaxies, such as accretion disc instabilities \citep{Saxton2018}, interacting supernovae (SNe; \citealt{Drake2011,Dong2016, Saxton2018}) or exotic stellar collisions \citep{Metzger2017}, several lines of evidence have emerged to suggest that at least the UV/optical TDEs candidates are due to the disruption of stars. These include i) their temperature and blackbody radius evolution, which is unlike any known SNe \citep{Hung2017, Holoien2018}, ii) their Eddington ratio and black hole mass distribution \citep{Wevers2017, Mockler2018}, iii) their luminosity function and volumetric rate as a function of black hole mass \citep{vanvelzen2018}, iv) their late-time UV emission, 5\,--\,10 years after peak brightness \citep{vanvelzen2018b}, and v) their location in galaxies without significant emission line content, indicating no on-going star formation nor AGN activity. For the X-ray selected candidates, \citet{Auchettl2018} studied the spectral and time evolution in comparison with an AGN sample and found that TDE candidates are significantly softer (see also e.g. \citealt{Lin2011}), less variable in terms of spectral hardness and display a more monotonic decay in their lightcurves.

Only a small fraction of UV/optical selected TDEs were observed to be X-ray bright, and similarly, many X-ray selected TDE candidates did not show contemporaneous UV/optical blackbody emission. For most of these events, however, this is explained by a lack of simultaneous observations and the true level of UV/optical emission is highly uncertain. Nevertheless, several objects have now been observed to be both optical and X-ray bright (e.g. ASASSN--14li, \citealt{Holoien2016}; ASASSN--15oi, \citealt{Holoien15oi, Gezari2017}; PS18kh, \citealt{Holoien2018, vanvelzen2018c}), suggesting that these flares all belong to the same class. The lack of X-ray emission in some UV/optical TDEs may be due to geometrical viewing angle effects, as in the AGN unification model \citep{Metzger2016, Dai2018}. The specific orbital dynamics of the events has also been suggested as the potential origin of the observational dichotomy. For example, \citet{Dai2015} suggested that X-ray emission may arise from more relativistic (i.e. deeper penetrating) encounters, while UV/optical emission may dominate in less relativistic TDEs. 

While the host galaxies of UV/optical selected TDEs have been extensively studied \citep{French2016, French2017, Hung2017, Law-Smith2017, Wevers2017, Graur2018}, the host galaxies of soft X-ray TDE candidates have received comparatively little attention. \citet{Graur2018} studied a sample of 35 TDE candidates (including both UV/optical and X-ray selected events) confirming the findings of \citet{Arcavi2014} and \citet{French2016} that, in the hosts of both TDE candidate classes there is an apparent overrepresentation of quiescent Balmer-strong (post-starburst or E+A) galaxies. In addition, they conclude that the large-scale properties (such as the density at the effective radius) of the hosts could be good predictors for their sub-pc scale properties. Interestingly, there exists a growing sample of optically discovered TDE candidates that were found in known AGN (e.g. \citealt{Merloni2015, Blanchard2017}). These events have well-studied host galaxies that were unambiguously identified as AGN based on their optical emission line content, whereas most of the X-ray selected TDE candidates have not been studied in such detail. These discoveries raise the question of whether a sample of such events is being missed by current time-domain surveys due to selection biases against spectroscopic follow-up of variability in known AGN host galaxies.

A recent study of 53000 galaxies in the Swift BAT archive by \citet{Walter2016} led to the discovery of a sample of hard X-ray flares in otherwise X-ray quiescent host galaxies. Based on the timescales and peak luminosities (of order 10$^{44}$ erg s$^{-1}$) that results from the association of some of these flares with the unique host galaxy within the X-ray error circle, several are unlikely to be AGN flares or Galactic in nature; instead, these events have characteristics consistent with the expectations for TDE candidates. However, these events have received little further attention in the literature due to the lack of multi-wavelength data for both the flares and hosts. One important caveat to the TDE interpretation is the poor spatial resolution of the BAT instrument. The typical localisation error circle is $\sim$\,2 arcmin, so these events were assumed to be associated to the only host galaxy within this error circle, with several offsets between the best-fit BAT position and the tentative host galaxy similar to the error circle size. On the other hand, given the low galaxy background density the probability of chance alignments is low, arguing in favour of the associations.\\

In this work we present medium resolution spectroscopic observations of a sample of TDE candidates selected by UV/optical, soft X-ray, and hard X-ray observations. We aim to characterise the nature of the host galaxies of the X-ray selected events, to discriminate between the AGN flare and TDE interpretations, based on the emission line content (or lack thereof) in the optical spectra. We constrain the black hole mass distribution with a total of 29 host galaxies (including both UV/optical and X-ray selected hosts) and discuss physical implications for the TDE properties. We present the observations in Section \ref{sec:observations}, and discuss the velocity dispersion measurements in Section \ref{sec:sigmas}. The new results of our work, including emission line ratios, black hole and host galaxy masses are discussed in Section \ref{sec:results}, and the implications of our measurements are explored in Section \ref{sec:discussion}. Our conclusions are presented in Section \ref{sec:conclusions}.

%%%%%%%%%%%%%%%%%%%%%%%%%%%%%%%%%%%%%%%%%%%%%%%%%%%%%%%%%%%%%%%%%%%%%%%
\section{Observations}
\label{sec:observations}
We obtained medium resolution optical long-slit spectra to characterise the emission line content, and measure the velocity dispersions and infer black hole masses for a sample of 17 X-ray selected TDE candidate host galaxies (Table \ref{tab:newobservations}). For the candidate TDE RXJ1242, we were able to obtain a spectrum only for RXJ1242A, the brighter of the two potential host galaxies. Our results for this source rely on the (currently unverified) assumption that this is indeed the correct host galaxy. In addition, we present new observations of 6 UV/optical selected TDEs. We also include archival observations from SDSS for the sources RBS 1032, SDSS J1323 and RX J1420 (this event has 2 potential host galaxies, but we follow \citealt{Graur2018} and assume RXJ1420A is the true host galaxy), and we use a measurement of the velocity dispersion from the Calar Alto Legacy Integral Field Area Survey (CALIFA) survey \citep{Garcia-Benito2015} for NGC 6021. Finally, we measure the velocity dispersion for ASASSN--15lh from the X-shooter spectrum presented in \citet{Kruhler2018}. Below we briefly describe the instrumental setup of the new WHT and Keck spectra and the data reduction process.

\begin{figure*}
  \includegraphics[width=\textwidth]{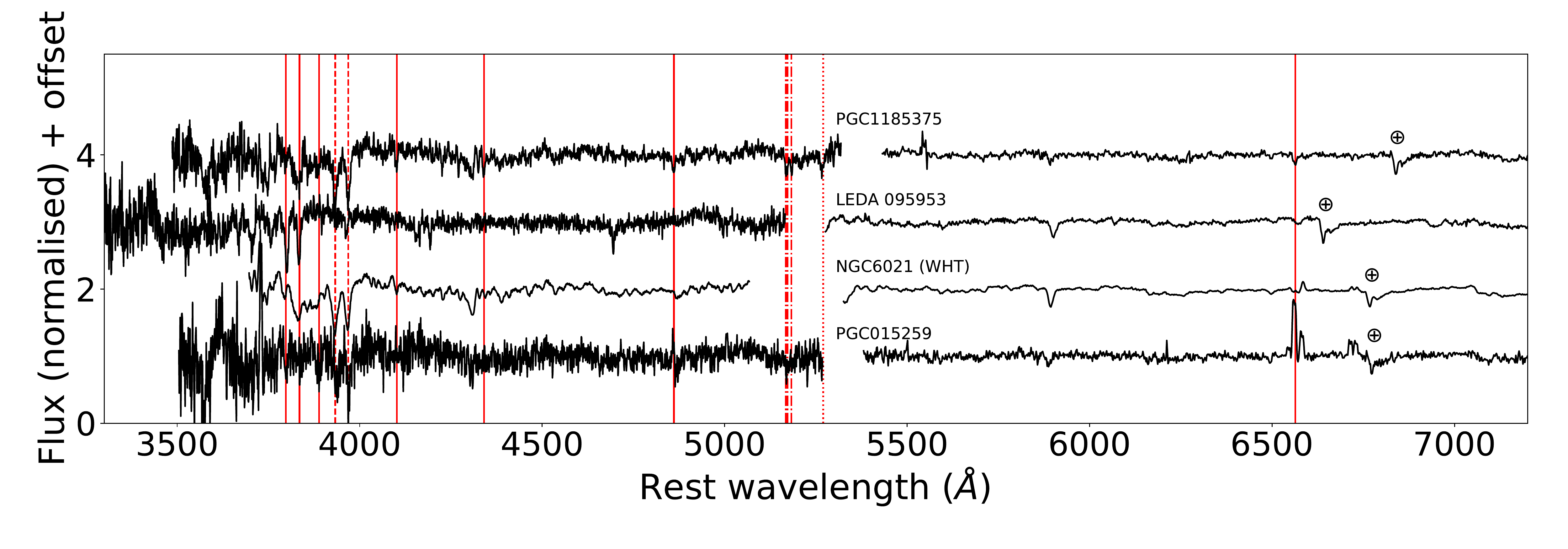}
  \includegraphics[width=\textwidth]{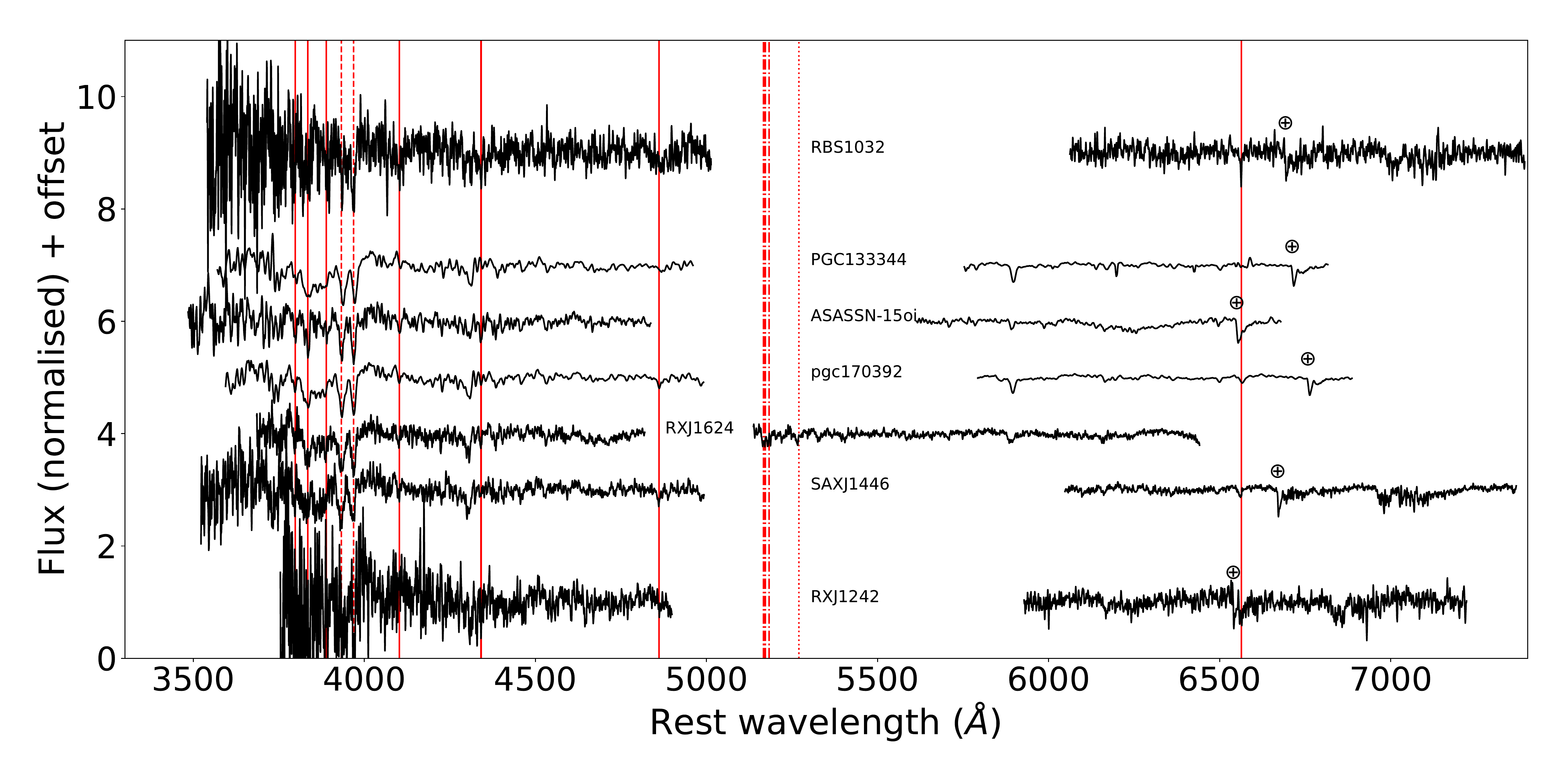}
  \caption{WHT/ISIS spectra of part of the sample of TDE hosts, in the host rest-frame. Top: the low resolution spectra taken with the R300B and R158R gratings. Bottom: the R600 spectra. The latter spectra have been smoothed by a boxcar filter of width 3 pixels for display purposes. The solid lines indicate the wavelengths of the Balmer series transitions, while dot-dashed lines show the Mg b triplet and dashed lines mark the Calcium H+K lines.}
\label{fig:whtspectra}
\end{figure*}

\begin{table*}
\centering
\caption{Overview of the spectroscopic observations used in this work. We note the instrument used for the spectra, which is either WHT ISIS, Keck ESI, VLT X-shooter, GTC OSIRIS, CAHA PMAS/PPak V1200 or SDSS. Slit denotes the slit width, and $\sigma_{instr}$ is the velocity dispersion resolution at 4000\,\AA\ as measured from skylines or arc lamp frames. For the soft X-ray events, the class is taken from \citet{Auchettl2017}. 2MASXJ1446 is a newly discovered X-ray TDE candidate (Saxton et al. in prep). The full observing log is given in Table \ref{tab:obslog}.}
  \begin{tabular}{lcccccc}
  \hline
  Name & RA & Decl. & Instrument & Slit/$\sigma_{instr}$ & z & Class   \\
   & hh mm ss.ss & dd mm ss.s & &arcsec / km s$^{-1}$  & & \\
  \hline
  2MASX J0249&  02 49 17.31 &--04 12 52.1&ESI & 0.5/16 & 0.019& Likely   \\
  3XMM J1500 & 15 00 52.07 & +01 54 53.8 & OSIRIS & 0.6/58 & 0.145 & Likely \\
  3XMM J1521 & 15 21 30.72 & +07 49 16.5 & OSIRIS & 0.6/58 & 0.179 & Likely \\
  LEDA 095953 & 13 47 29.79 & --32 54 51.9 & ISIS & 1.5/400 & 0.035 & Possible \\
  2MASX J1446 & 14 46 05.22 & +68 57 31.1 & ISIS & 0.7/45 & 0.030 & X-ray \\ 
  NGC 5905 & 15 15 23.48 &+55 31 05.9&ISIS & 1.0/59 & 0.011& Possible / AGN?    \\
  RBS 1032 &11 47 26.69  &+49 42 57.7 & SDSS & fiber& 0.026 & Possible  \\ 
  RX J1242A & 12 42 38.55 &--11 19 20.8 & ISIS& 0.6/40 & 0.050 &Possible  \\ 
  RX J1420A & 14 20 24.37 & +53 34 11.7 &SDSS & fiber & 0.147 &Possible  \\
  RX J1624& 16 24 57.18 &+75 54 54.3&ISIS &1.1/66 & 0.064 & Possible \\
  SDSS J1201 & 12 01 36.03 & +30 03 05.5& ESI& 0.5/16 & 0.146 & Likely  \\
  SDSS J1323 & 13 23 41.97 & +48 27 01.3 & SDSS & fiber & 0.088 & Likely  \\
  SDSS J0159 &01 59 57.64 & +00 33 10.5& ESI& 0.5/16 &  0.311 &Possible / CLAGN?  \\\hline 
  NGC 6021 & 15 57 30.68 &+15 57 22.4& ISIS & 1.5/400 &  0.016 & Hard X-ray  \\
  NGC 6021 &  15 57 30.68 &+15 57 22.4& V1200 & fiber & 0.016& Hard X-ray  \\ 
  PGC 015259 & 04 29 21.82 &--04 45 35.7 & ISIS & 1.5/400 & 0.015 & Hard X-ray  \\ 
  PGC 1127937 &01 18 56.60 & --01 03 10.8 & ESI& 0.5/16 & 0.020 & Hard X-ray \\
  PGC 1185375 & 15 03 50.29 & +01 07 36.7 & SDSS & fiber& 0.005 & Hard X-ray  \\  
  PGC 133344 & 21 42 56.03& --30 08 00.3&ISIS & 1.0/59 &0.024 & Hard X-ray \\
  PGC 170392 & 22 26 46.22 & --15 01 18.6 & ISIS & 1.0/59 &  0.016&Hard X-ray \\
  UGC 1791& 02 19 53.66 & +28 14 52.6& ESI& 0.5/16 &  0.016&Hard X-ray  \\\hline
  ASASSN15--lh &22 02 15.39 & --61 39 34.6 & X-Shooter & 1.0/55 & 0.225& Optical / SN?  \\
  ASASSN15--oi &20 39 09.03 &--30 45 20.8 &ISIS & 1.0/59 & 0.048 & Optical\,+\,X-ray  \\
  DES14C1kia & 03 34 47.49 & --26 19 35.0&ESI& 0.5/16 &  0.162 & Optical \\
  GALEX D1--9 & 02 25 17.00 & --04 32 59.0 & ESI & 0.5/16 &  0.326 & Optical \\
  GALEX D23H1 & 23 31 59.54 & +00 17 14.6  & ESI& 0.5/16 &  0.186 & Optical  \\
    PS1--11af & 09 57 26.82 & 03 14 00.9 & ESI & 0.5/16 &  0.405 & Optical \\
  SDSS TDE2 & 23 23 48.62 & --01 08 10.3 & ESI& 0.5/16 &  0.252 & Optical  \\\hline
    \end{tabular}
  \label{tab:newobservations}
\end{table*}

\begin{table*}
\centering
\caption{Measured properties of the sources used in this work. $\sigma$ is the velocity dispersion (where an asterisk indicates the use of a fiber or galaxy wide extraction), and M$_{\rm BH}$ the derived black hole mass. M$_{\rm g}$ and log(M$_*$) present the absolute g-band magnitude and total stellar mass of the host, respectively. These values are taken from \citet{vanvelzen2018} for the optical hosts, while for the X-ray hosts stellar masses are computed from SDSS photometry if available, and from PS1 photometry otherwise. BPT lists the classification in the BPT diagram for galaxies with detectable emission lines. Here Q stands for quiescent, Sy for Seyfert, SF for star-forming, C for composite SF+AGN and L for LINER. L$_{\rm max}$ provides the maximum {\it observed} X-ray luminosity for the X-ray events, while for the optical events the integrated blackbody luminosity at peak is given. R$_{\rm BB}$ provides the blackbody radius derived from L$_{\rm max}$, and R$_{\rm petro}$ is the 90\,$\%$ light radius taken from SDSS. The sources below the double horizontal line are the optical TDEs presented in \citet{Wevers2017}. This Table is available in machine-readable form.}
  \begin{tabular}{lccccccccc}
  \hline
  Name & $\sigma$ & log(M$_{\rm BH}$)  &  M$_{\rm g}$ & log(M$_*$) & BPT & L$_{\rm max}$ & R$_{\rm BB}$ & R$_{\rm petro}$ & Notes \\\vspace{0.5mm}
   & [km s$^{-1}$] & [M$_{\odot}$] & [mag] & [M$_{\odot}$] & &[erg s$^{-1}$] & [cm] & [arcsec]& \\
  \hline\vspace{0.5mm}
    2MASX J0249&  43\,$\pm$\,4 & 4.93$^{+0.55}_{-0.53}$ & --17.5 & 9.1 & SF/C &  3.4\,$^{+3.6}_{-3.0}$\,$\times$\,10$^{41}$ & 2.9$^{+3.3}_{-3.1}$\,$\times$\,10$^{10}$ & 5.4 &  \\\vspace{0.5mm}
 3XMM J1500 & 59\,$\pm$\,3$^*$ & 5.64$^{+0.45}_{-0.45}$& --19.1& 9.3& SF & 6.2\,$^{+1.6}_{-1.3}$\,$\times$\,10$^{43}$& 3.9$^{+3.9}_{-3.9}$\,$\times$\,10$^{11}$ & 1.8 & \citet{Lin2017}\\\vspace{0.5mm}
 3XMM J1521 & 58\,$\pm$\,2 & 5.61$^{+0.41}_{-0.41}$& --19.2& 9.9 & Q & 3.2\,$^{+3.5}_{-2.9}$\,$\times$\,10$^{43}$& 2.8$^{+3.2}_{-3.1}$\,$\times$\,10$^{11}$ & 1.7 & \\\vspace{0.5mm}
    LEDA 095953 & --- & --- & --- & --- & Q & 5.4\,$^{+5.8}_{-5.2}$\,$\times$\,10$^{42}$ & ---& --- & mQBS \\\vspace{0.5mm}
    2MASX J1446 & 167\,$\pm$\,15 & 7.84$^{+0.54}_{-0.52}$ & --19.6 & 9.8 &Q & 4\,$^{+2}_{-2}$\,$\times$\,10$^{42}$ & 9.9$^{+10}_{-10}$\,$\times$\,10$^{10}$& --- & \\\vspace{0.5mm} 
    NGC 5905 & 97\,$\pm$\,5 & 6.69$^{+0.45}_{-0.44}$ &--20.2  &  10.0  &  SF/C & 8.7\,$^{+9.3}_{-8.1}$\,$\times$\,10$^{40}$ & 1.5$^{+1.7}_{-1.6}$\,$\times$\,10$^{10}$& 45.8 & AGN?  \\\vspace{0.5mm}
    RBS 1032 & 49\,$\pm$\,7$^*$ & 5.25$^{+0.67}_{-0.62}$ & --17.7 & 9.0& Q&5.0\,$^{+5.5}_{-4.4}$\,$\times$\,10$^{41}$ & 3.5$^{+4.0}_{-3.8}$\,$\times$\,10$^{10}$& 4.0 & \\\vspace{0.5mm} 
    RX J1242A & --- & --- & --20.5 & 10.3 & Q & 4.0\,$^{+4.6}_{-2.8}$\,$\times$\,10$^{42}$ & ---& 11.3 &\\ 
    RX J1420A & 131\,$\pm$\,13$^*$ & 7.33$^{+0.56}_{-0.54}$ & --20.3 & 10.3 & Q& 2.4\,$^{+2.6}_{-2.1}$\,$\times$\,10$^{43}$ & 2.4$^{+2.7}_{-2.6}$\,$\times$\,10$^{11}$ & 3.7 & \\
    RX J1624 & 155\,$\pm$\,9 & 7.68$^{+0.45}_{-0.45}$ & --20.8 & 10.4  & Q  & 2.4\,$^{+2.5}_{-2.3}$\,$\times$\,10$^{43}$ &2.4$^{+2.7}_{-2.7}$\,$\times$\,10$^{11}$ & --- & \\\vspace{0.5mm}
  SDSS J0159 & 124\,$\pm$\,10 & 7.21$^{+0.52}_{-0.50}$ & --21.8 & 10.7 & C & 1.1\,$^{+1.1}_{-1.0}$\,$\times$\,10$^{44}$ &5.2$^{+5.8}_{-5.7}$\,$\times$\,10$^{11}$ & 1.7 & CL AGN?  \\\vspace{0.5mm}
  SDSS J1201 & 122\,$\pm$\,4  & 7.18$^{+0.41}_{-0.41}$ & --20.6 & 10.4 & Q &  1.0\,$^{+1.2}_{-0.8}$\,$\times$\,10$^{45}$ & 1.6$^{+1.8}_{-1.7}$\,$\times$\,10$^{12}$ &4.2 &  \\\vspace{0.5mm}
  SDSS J1323 & 75\,$\pm$\,4$^*$ & 6.15$^{+0.46}_{-0.45}$ & --18.9 & 9.8 &Q& 2.0\,$^{+2.4}_{-1.6}$\,$\times$\,10$^{44}$ & 7.0$^{+8.2}_{-7.5}$\,$\times$\,10$^{11}$ &4.8&  \\\hline\vspace{0.5mm} 
  PGC 015259 & --- & --- & --18.6 & 9.5 & SF/C  & 1.6\,$\times$\,10$^{44}$ & ---& --- &\\\vspace{0.5mm}  
  NGC 6021 &  187\,$\pm$\,3& 8.08$^{+0.37}_{-0.37}$& --20.4 &10.3 &  Sy&1.8\,$\times$\,10$^{44}$ & --- & --- &AGN\\\vspace{0.5mm}
  PGC 1127938 & 31\,$\pm$\,2$^*$ & 4.29$^{+0.55}_{-0.54}$ & --16.4 & 8.6 & Q & 2.5\,$\times$\,10$^{44}$ &7.8$^{+8.1}_{-8.1}$\,$\times$\,10$^{11}$ & 8.6 &\\\vspace{0.5mm} 
  PGC 1185375 & 41\,$\pm$\,7$^*$ & 4.86$^{+0.46}_{-0.45}$  & --16.1 & 8.5 &Q& 4.2\,$\times$\,10$^{43}$ & 3.2$^{+3.3}_{-3.3}$\,$\times$\,10$^{11}$ & 20.3 & \\\vspace{0.5mm} 
  PGC 133344 & 173\,$\pm$\,3& 7.91$^{+0.38}_{-0.38}$ & --19.9& 10.1  & Q & 3.5\,$\times$\,10$^{44}$ & 9.3$^{+9.5}_{-9.5}$\,$\times$\,10$^{11}$&---& \\\vspace{0.5mm}
  PGC 170392 & 169\,$\pm$\,3 & 7.86$^{+0.38}_{-0.38}$ & --- &---  & Q& 2.8\,$\times$\,10$^{44}$ & 8.3$^{+8.5}_{-8.5}$\,$\times$\,10$^{11}$& ---&\\\vspace{0.5mm}
  UGC 1791& 41\,$\pm$\,4$^*$ & 4.86$^{+0.56}_{-0.54}$ &--16.7 & 8.1 & SF &  3.0\,$\times$\,10$^{44}$ & 8.6$^{+8.8}_{-8.8}$\,$\times$\,10$^{11}$&---&\\\hline\vspace{0.5mm}
    ASASSN15--lh &210\,$\pm$\,7 & 8.32$^{+0.41}_{-0.41}$ & --21.4 &10.8  & L &  4.2\,$^{+1.0}_{-0.9}$\,$\times$\,10$^{45}$ & 3.9$^{+0.9}_{-0.9}$\,$\times$\,10$^{15}$& ---&\\\vspace{0.5mm}
    ASASSN15--oi &61\,$\pm$\,7 & 5.71$^{+0.60}_{-0.57}$ & --19.3&9.9 & Q& 2.8\,$^{+0.7}_{-0.6}$\,$\times$\,10$^{44}$ &1.0$^{+0.2}_{-0.2}$\,$\times$\,10$^{15}$ &---& L$_{\rm X}$\,=\,3.1\,$^{+3.7}_{-2.7}$\,$\times$\,10$^{41}$ \\\vspace{0.5mm}
DES14C1kia & ---&--- &--20.1 & 10.1 & Q & --- &  ---&---&mQBS?  \\\vspace{0.5mm}
    GALEX D1--9 & 65\,$\pm$\,6 & 5.85$^{+0.54}_{-0.53}$ & --20.0 & 10.3 & Q & 1.2\,$^{+0.3}_{-0.3}$\,$\times$\,10$^{44}$ & 1.3$^{+0.3}_{-0.3}$\,$\times$\,10$^{14}$ & 2.8& \\\vspace{0.5mm}
    GALEX D23--H1 & 84\,$\pm$\,4$^*$ & 6.39$^{+0.44}_{-0.44}$ & --20.1& 10.3 & SF & 9\,$^{+3}_{-2}$\,$\times$\,10$^{43}$ & 1.5$^{+0.4}_{-0.4}$\,$\times$\,10$^{14}$& 2.1&\\\vspace{0.5mm}
    PS1--11af & --- & --- & --20.1 & 10.1 & Q & 7.2\,$^{+0.8}_{-0.7}$\,$\times$\,10$^{43}$ & --- &1.3& \\\vspace{0.5mm}
    SDSS TDE2 & --- & ---& --20.6&10.6  & Q & 1.0\,$^{+0.1}_{-0.1}$\,$\times$\,10$^{44}$ &---& 3.2& \\\hline\hline\vspace{0.5mm}
  ASASSN--14li & 81\,$\pm$\,2 & 6.23$^{+0.39}_{-0.40}$ & --18.8 & 9.6 & Q & 6.2\,$^{+1.4}_{-1.2}$\,$\times$\,10$^{43}$ & 2.4$^{+0.5}_{-0.5}$\,$\times$\,10$^{14}$ & 3.1&L$_{\rm X}$\,=\,1.1\,$^{+1.1}_{-1.0}$\,$\times$\,10$^{43}$ \\\vspace{0.5mm}
  ASASSN--14ae & 53\,$\pm$\,2 & 5.42$^{+0.46}_{-0.46}$ & --19.2 & 9.8 & Q & 7.5\,$^{+1.6}_{-1.4}$\,$\times$\,10$^{43}$ & 7$^{+1.5}_{-1.5}$\,$\times$\,10$^{14}$ &5.0& \\\vspace{0.5mm}
  GALEX D3--13 & 133\,$\pm$\,6$^{*}$ & 7.36$^{+0.43}_{-0.44}$ &--20.8 & 10.7 & Q & 2.0\,$^{+0.2}_{-0.2}$\,$\times$\,10$^{44}$ & 2.2$^{+0.2}_{-0.2}$\,$\times$\,10$^{14}$ & 1.7&\\\vspace{0.5mm}
  PTF--09ge & 82\,$\pm$\,2  & 6.31$^{+0.39}_{-0.39}$ &--19.5 & 10.1 & Q & 1.3\,$^{+0.3}_{-0.3}$\,$\times$\,10$^{44}$ & 9$^{+2}_{-2}$\,$\times$\,10$^{14}$ & 5.5&\\\vspace{0.5mm}
  PTF--09axc & 60\,$\pm$\,4 & 5.68$^{+0.48}_{-0.49}$ &--20.2 & 10.0 & Q & 3.1\,$^{+0.4}_{-0.4}$\,$\times$\,10$^{43}$ & 1.45$^{+0.03}_{-0.03}$\,$\times$\,10$^{15}$ & 2.5&\\\vspace{0.5mm}
  PTF--09djl & 64\,$\pm$\,7 & 5.82$^{+0.56}_{-0.58}$ &--19.6 & 10.1 & Q & 2.5\,$^{+0.6}_{-0.5}$\,$\times$\,10$^{44}$ &  9$^{+2}_{-2}$\,$\times$\,10$^{14}$&2.0& \\\vspace{0.5mm}
  iPTF--15af & 106\,$\pm$\,2 & 6.88$^{+0.38}_{-0.38}$ &--17.9 & 10.2 & Q & 1.2\,$^{+1.1}_{-0.5}$\,$\times$\,10$^{44}$ & 2.0$^{+1.3}_{-1.3}$\,$\times$\,10$^{14}$&4.5& \\\vspace{0.5mm}
  iPTF--16axa & 82\,$\pm$\,3 & 6.34$^{+0.42}_{-0.42}$ & --19.4 & 10.1 & Q & 3.3\,$^{+0.9}_{-0.7}$\,$\times$\,10$^{44}$ & 7.6$^{+1.5}_{-1.5}$\,$\times$\,10$^{14}$&3.2& mQBS\\\vspace{0.5mm}
  iPTF--16fnl & 55\,$\pm$\,2 & 5.50$^{+0.42}_{-0.42}$ & --19.8 & 9.8 & Q & 3.3\,$^{+0.9}_{-0.7}$\,$\times$\,10$^{43}$ & 1.8$^{+0.4}_{-0.4}$\,$\times$\,10$^{14}$ &10.4& \\\vspace{0.5mm}
  PS1--10jh & 65\,$\pm$\,3 & 5.85$^{+0.44}_{-0.44}$ & --18.1 & 9.5 & Q & 1.6\,$^{+0.3}_{-0.3}$\,$\times$\,10$^{43}$ & 5.7$^{+0.9}_{-0.9}$\,$\times$\,10$^{14}$& 1.8&\\\vspace{0.5mm}
  SDSS TDE1 & 126\,$\pm$\,7 & 7.25$^{+0.45}_{-0.46}$ & --19.2 & 10.1 & Q & 3.0\,$^{+1.0}_{-0.8}$\,$\times$\,10$^{43}$ & 3.6$^{+0.1}_{-0.1}$\,$\times$\,10$^{14}$ & 2.1&\\\hline
    \end{tabular}
  \label{tab:results}
\end{table*}

\subsection{WHT/ISIS}
\label{sec:wht}
Part of the observations were performed using the Intermediate dispersion Spectrograph and Imaging System (ISIS, \citeauthor{Jorden1990} \citeyear{Jorden1990}) mounted at the Cassegrain focus of the 4.2m William Herschel Telescope (WHT) situated on the Canary island of La Palma, Spain. Typically, we obtained spectra with the R600 gratings in both arms, in combination with the dichroic at 5300\,\AA. Some sources were observed with the R158R and R300B gratings. These latter are of too low spectral resolution to measure velocity dispersions below 400 km s$^{-1}$, but they do allow us to measure the emission line content. We ensured that the R600 grating observations were performed in slit-limited observing conditions, such that the instrumental resolution can be measured from sky emission lines or arc lamp observations if no sky lines are present. The spectra are presented in Figure \ref{fig:whtspectra}.

We subtract the bias level, perform a flatfield correction and finally apply a wavelength calibration using CuNe+CuAr arc lamp frames in \textsc{iraf}. Cosmic rays are removed using the \textit{lacos} package in \textsc{iraf} \citep{vanDokkum2012}. We perform an optimal extraction of the spectra \citep{Horne1986} using an aperture with a size in the spatial direction equal to the slit width to obtain a spectrum of the central region of the galaxy. We also rebin the spectra to a linear dispersion on a logarithmic wavelength scale. From the associated arc lamp observations and/or sky emission lines we measure the instrumental broadening, which is typically a FWHM resolution of 138\,$\pm$\,1 \kms\ at 4000\,\AA\ for a 1\arcsec\ slit and the blue R600 grating, but the actual value depends (linearly) on the slit width (see Table \ref{tab:newobservations}). This correponds to a velocity dispersion of 59 km s$^{-1}$.

\subsection{Keck/ESI}
\label{sec:keck}
\begin{figure*}
  \includegraphics[width=\textwidth]{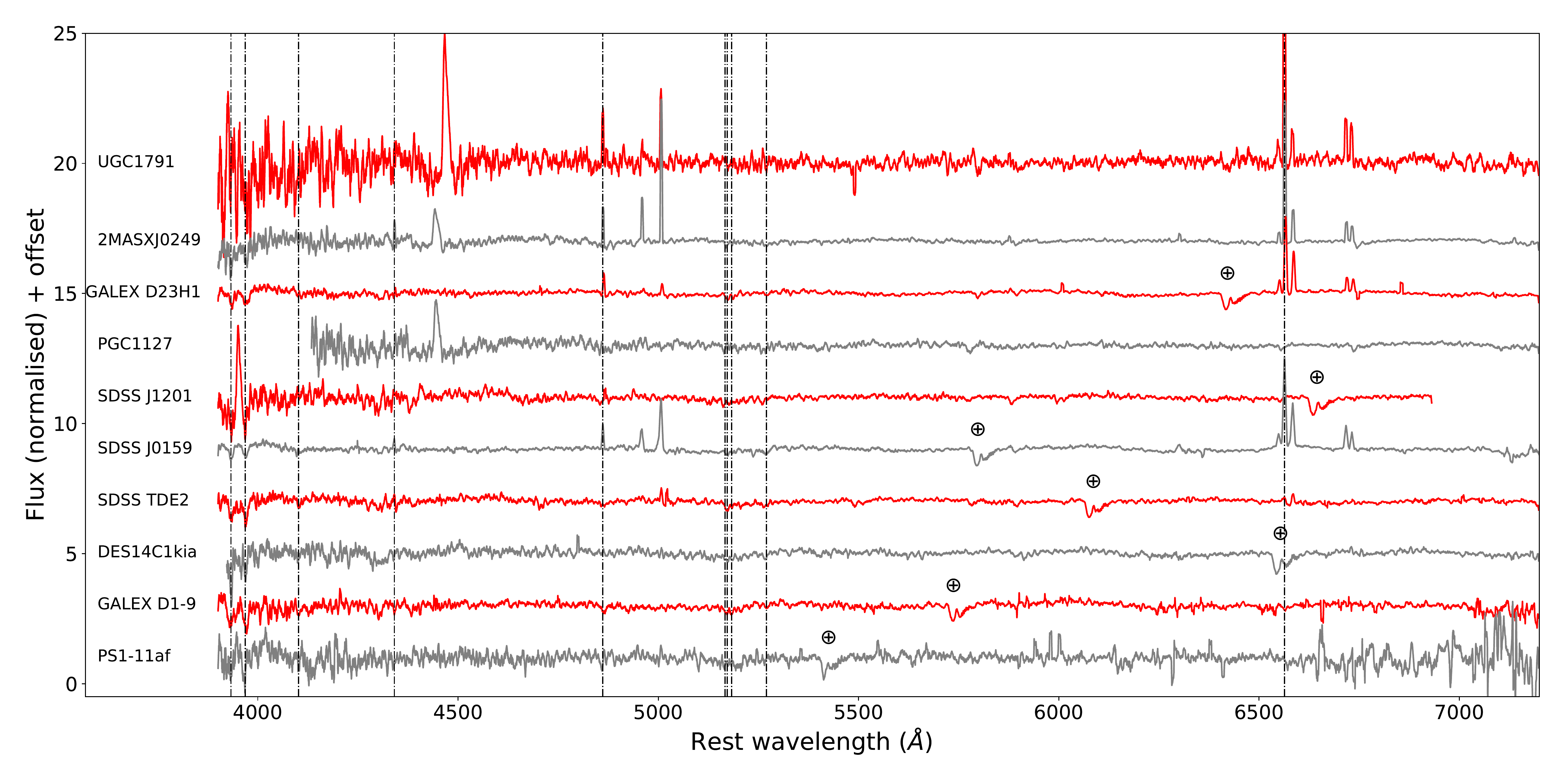}
  \caption{Keck/ESI spectra of part of the sample of TDE hosts, in the host rest-frame. The spectra have been smoothed by a boxcar filter of width 25 pixels for display purposes. There is a broad feature near 4500\,\AA\ (in the observed frame) that is caused by a detector artefact. The vertical lines are as in Figure \ref{fig:whtspectra}.}
\label{fig:keckspectra}
\end{figure*}

We also obtained medium resolution spectra using the Echelette Spectrograph and Imager (ESI; \citealt{Sheinis2002}), mounted at the Cassegrain focus of the Keck--II telescope on Mauna Kea, Hawaii. The spectra were taken on 2017 November 17, using a 0\farcs5 slit. This setup delivers a near-constant resolving power of R\,=\,8000, corresponding to a FWHM resolution of 38 km s$^{-1}$ at 4000\,\AA\ (or $\sigma$\,=\,16 km s$^{-1}$). The spectra are presented in Figure \ref{fig:keckspectra}.

The data were reduced using the MAuna Kea Echelle Extraction (\texttt{makee}) software package. The pipeline performs standard spectroscopic data reduction routines including a bias subtraction, a flatfield correction and spectrum extraction. A spectrum of a spectrophotometric standard star on the CCD was used to determine the trace of the science objects. The position of each echelle order is traced, optimally extracted and wavelength calibrated independently (using CuAr and HgNe+Xe arc lamp exposures), after which the orders are rebinned to a linear dispersion on a logarithmic wavelength scale with a constant dispersion of 11.5 km s$^{-1}$ per pixel and combined using the {\it combine} command. 

\subsection{GTC/OSIRIS}
\begin{figure*}
  \includegraphics[width=\textwidth]{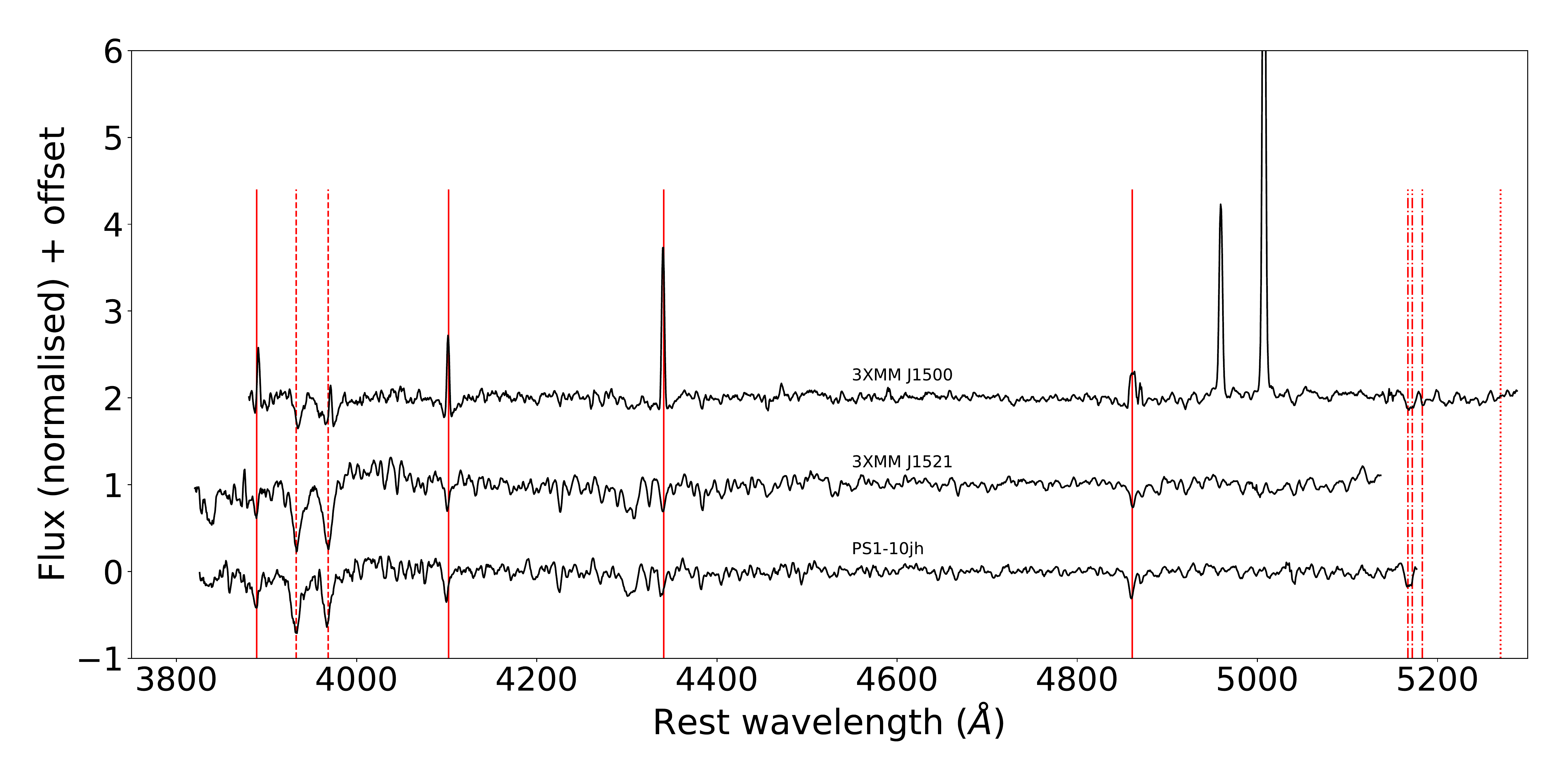}
  \caption{GTC/OSIRIS spectra of part of the sample of TDE hosts, in the host rest-frame. The spectra have been smoothed by a boxcar filter of width 5 pixels for display purposes. The vertical lines are as in Figure \ref{fig:whtspectra}.}
\label{fig:gtcspectra}
\end{figure*}

Three sources were observed with the 10.4m Gran Telescopio Canarias (GTC) located on La Palma, Spain using the Optical System for Imaging and low Resolution Integrated Spectroscopy (OSIRIS, \citealt{Cepa2000}) instrument operated in long-slit spectroscopy mode. These data have been reduced using a custom \textsc{python} semi-automatic routine, based on \textsc{iraf} and \textsc{molly}\footnote{\textsc{molly} is software developed by T. R. Marsh for the reduction and analysis of spectroscopic data.} tasks. The spectra are first bias subtracted and flatfield corrected. Cosmic rays are removed using the \textit{lacos} package in \textsc{iraf}, and we use an optimal extraction with an aperture of 0.6 arcsec. Individual arc spectra are extracted from the two-dimensional images at the position defined by the continuum trace of the science target. We measure a FWHM of 138\,$\pm$\,2 km s$^{-1}$ for the 0\farcs6 slit width, corresponding to a velocity dispersion of 59 km s$^{-1}$ at 4000 \AA. To derive a precise wavelength calibration, we interpolate the results from arcs obtained before and after the observation; when not available, the nearest arc is selected. This wavelength calibration is further refined by comparing the sky emission line O\,\textsc{i} 5577.3 \AA\ with its corresponding rest wavelength, from which we derive sub-pixel velocity drifts ($<$ 20 km/s) that are subsequently corrected. Finally, we remove the Earth velocity relative to the source at each observational epoch to obtain the final spectra in the solar system barycenter reference frame. The spectra are shown in Figure \ref{fig:gtcspectra}.\\

All spectra are normalised to the continuum by fitting 3$^{\rm rd}$ order cubic splines in \textsc{molly}. We mask prominent absorption and emission lines during this process. Spectra for which multiple exposures were obtained are subsequently averaged by weighting each exposure with its mean signal-to-noise ratio (variance). We provide the full observing logs, including dates and exposure times, in Table \ref{tab:obslog}. The spectra can be made available upon request to the authors.

\section{Velocity dispersion measurements}
\label{sec:sigmas}
In this work we use the Penalized Pixel Fitting (\textsc{ppxf}) routine \citep{Cappellari2017} in combination with the Elodie stellar template library (985 templates) to measure the stellar velocity dispersion using the absorption lines present in the spectra. We refer the reader to \citet{Wevers2017} for a detailed discussion of the method and caveats. Briefly, the set of 985 templates is compared to the galaxy spectrum, after which a limited number (typically 10\,--\,20) of templates is chosen to perform a detailed fit to the velocity (redshift) and line-of-sight velocity dispersion (LOSVD) of the spectrum. To this end, the templates are convolved with the LOSVD, which is parametrised by a series of Gauss-Hermite polynomials (up to 4$^{\rm th}$ order), after which the best-fit template is determined by $\chi^2$ minimisation (see \citealt{Cappellari2017} for a detailed explanation of \textsc{ppxf}). 

\begin{figure*}
 \includegraphics[width=\textwidth]{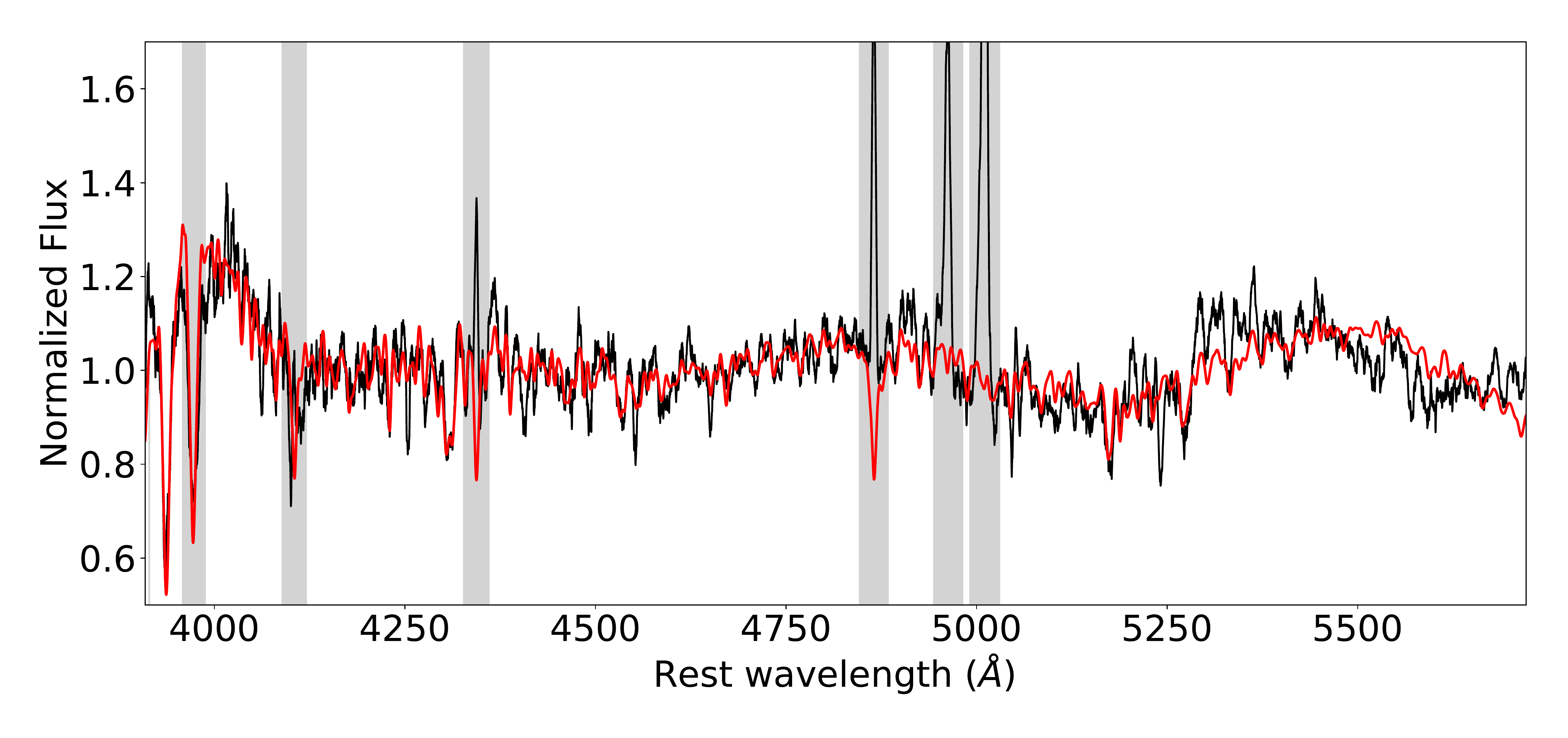}
 \caption{Best-fit template (red), broadened to a velocity dispersion of 124 km s$^{-1}$, overplotted on the (smoothed) Keck spectrum of SDSS J0159 (black). The shaded regions are excluded during the fitting process.}
\label{fig:examplefit}
\end{figure*}

While we always aim to use only the central region of the galaxy spectrum (using, as outlined in Section \ref{sec:wht}, an extraction box with side equal to the slit width), some spectra have low SNR, leading to degeneracies in the fitting routine. In this case, we use instead an extraction region which covers the entire galaxy along the length of the slit in order to increase the SNR. During fitting, we mask the H Balmer lines (because they are known to be strongly Stark broadened), as well as emission lines of [O\,$\textsc{iii}$] at $\lambda\lambda$4959, 5007, the diffuse interstellar band at 5780\,\AA\ and the Na\,$\textsc{i}$ interstellar absorption lines at $\lambda\lambda$\,5890, 5895. We show an example fit to the spectrum of SDSS J0159 in Figure \ref{fig:examplefit}, where the (smoothed) Keck spectrum is shown in black and the best-fit template broadened to 124 km s$^{-1}$ is shown in red. The shaded regions are exluded during the fitting process.
In some cases, the preliminary template fitting step in \textsc{ppxf} selects templates that do not represent the observed galaxy spectrum well, likely as a result of the limited SNR. As a consequence, the resulting velocity dispersion distribution does not converge to a consistent value. If this is the case, we limit the initial template library to a subset of 24 stellar templates, chosen to adequately cover the spectral range between A0 and M0, to decrease the degeneracy of input templates. We have verified that this yields unbiased estimates of the velocity dispersion for the Keck spectra that have robust measurements using the full template library. 

Uncertainties are estimated by perturbing the spectrum fluxes within the errors (assuming these follow a Gaussian distribution). We use the same procedure of template fitting to obtain a distribution of velocity dispersion measurements for each spectrum, containing at least 1000 samples. We fit this distribution with a Gaussian, and adopt the mean as the velocity dispersion and the standard deviation as the measurement uncertainty. 

Finally, we note that there are spectra, in particular the host galaxies of DES14C1kia, PS1--11af and TDE2, for which visual inspection showed that the best matching templates are a poor fit, i.e. the templates do not provide an adequate representation of the observed galaxy spectra. As such, the velocity dispersion values we measure are deemed to be inaccurate and unreliable, and we do not infer black hole masses for these galaxies. This may be related to the relatively small R$_{\rm petro}$ (as a proxy for galaxy size), indicating that our slit contains a large fraction of the host galaxy beyond the central bulge. As this is also the case for D1--9, we perform a consistency check using the galaxy extraction and find a significantly different result ($\sigma$\,=\,89\,$\pm$\,4 km s$^{-1}$) compared to the central extraction, indicating that the systematic errors for this source may be larger than the uncertainties quoted in Table \ref{tab:results}. For 3XMM J1500, we find a small difference between a narrow aperture extraction (45\,$\pm$\,4 km s$^{-1}$) and a wider extraction (59\,$\pm$\,3 km s$^{-1}$), consistent within the 3$\sigma$ measurement uncertainties. We adopt the latter value because this spectrum has a higher SNR. Furthermore for three sources, PGC 1127938, PGC 1185375 and UGC 1791, the templates provide a poor fit to the overall galaxy spectrum but it is clear that the absorption lines are narrow. For these galaxies we indeed measure very low velocity dispersion values ($\leq$\,40 km s$^{-1}$). Although the systematic errors may be larger than the statistical errors quoted in Table \ref{tab:results}, we do not exclude these measurements as there is additional evidence that the black holes inhabiting these galaxies must be small ($\leq$\,10$^6$\,M$_{\odot}$, see Section \ref{sec:comparison}). 

We use the measured velocity dispersions to infer the black hole mass using the M\,--\,$\sigma$ relation presented in \citet{Ferrarese2005}. We remark that especially at the low end (velocity dispersions $\leq$\,70 km s$^{-1}$), it is unclear (both theoretically and observationally) whether galaxies should still follow the same scaling relations as derived for massive ellipticals (e.g. \citealt{Graham2008, Volonteri2009, Xiao2011}). Given that this debate is ongoing, we will assume for now that indeed the same scaling laws are applicable throughout the mass range considered here ($\sim$\,10$^5$\,--\,10$^8$\,M$_{\odot}$). 

\section{Results}
\label{sec:results}
In this Section we will first discuss new and potential moderately quiescent Balmer strong (mQBS; e.g. \citealt{Graur2018}) hosts, after which we analyse the emission line content of the galaxies, finding that the majority of both optical and X-ray host galaxies do not show evidence for significant emission lines. We end by presenting the black hole mass distribution of all the galaxies in our sample, as well as galaxy stellar mass and bulge masses.

\subsection{Quiescent Balmer strong galaxies}
As first noted by \citet{Arcavi2014} for UV/optical selected TDEs, and later confirmed by \citet{French2016} and \citet{Graur2018}, TDEs are found to be overrepresented in quiescent, moderately Balmer strong galaxies (i.e. galaxies that experienced a recent starburst or have a truncated star formation history). We adopt the definition of \citet{Graur2018} based on the Lick absorption index that a moderately Balmer strong galaxy should have an H$\delta_{A}$ equivalent width (EW)\,$\geq$\,1.31\,\AA\ in absorption, while the H$\alpha$ EW should be less than 3\,\AA\ in emission. 
We find one source that satisfies these criteria. For the host galaxy LEDA 095953, we measure H$\delta_{A}$ EW\,=\,1.8\,\AA\ and H$\alpha$ EW\,=\,1.9\,\AA\ in absorption. We estimate that the uncertainties on these measurements are $\sim$\,0.5\,\AA. To obtain a lower limit on the H$\alpha$ EW, we correct the measurement by 2.5\,\AA\ to estimate the EW corrected for stellar absorption. This is motivated by the finding of \citet{French2016} that the largest stellar absorption correction to the H$\alpha$ EW for sources in their sample is $\sim$\,2.5\,\AA. This yields H$\alpha$ EW\,$\geq$\,--0.4\,\AA\ for LEDA 095953. In addition, we measure EW(H$\delta_{A}$)\,=\,3.5\,\AA\ for the host of DES14C1kia. Unfortunately, the H$\alpha$ line is redshifted into the strong atmospheric absorption band near 7800\,\AA, making it difficult to gauge whether there is absorption or emission in the H$\alpha$ line. We do not identify any narrow Balmer emission lines in the spectrum of DES14C1kia, nor any forbidden narrow emission lines, such as the Si or N doublets, nor O\,\textsc{iii} at 5007\,\AA, that would indicate ongoing star formation or nuclear black hole activity. This host galaxy could therefore be an additional member of the (m)QBS host galaxy population, although the lack of significant H$\alpha$ emission would require a spectrum that does not suffer from atmospheric absorption bands to confirm.

Additionally, while remeasuring the Lick index for the source iPTF--16axa, we find a value of EW(H$\delta_{A}$)\,=\,4.1\,\AA, inconsistent with the measurement of \citet{Graur2018}. Our measured value would result in a classification as mQBS rather than quiescent, even when adopting the same measurement uncertainty of 1.5\,\AA. This difference can potentially be attributed to the fact that in \citet{Graur2018}, an extraction of the whole galaxy light along the slit was used to allow for a fair comparison to the wide fiber measurements from SDSS, whereas we repeated the measurement using the central extraction. Gradients in the H$\delta_{A}$ EW as a function of distance to the nucleus have been observed in other post starburst galaxies \citep{Pracy2012}. We therefore tentatively classify the host of iPTF--16axa as an mQBS galaxy. 

\subsection{Emission line ratios: BPT diagrams}
\label{sec:bpt}
\begin{figure*}
 \includegraphics[width=\textwidth]{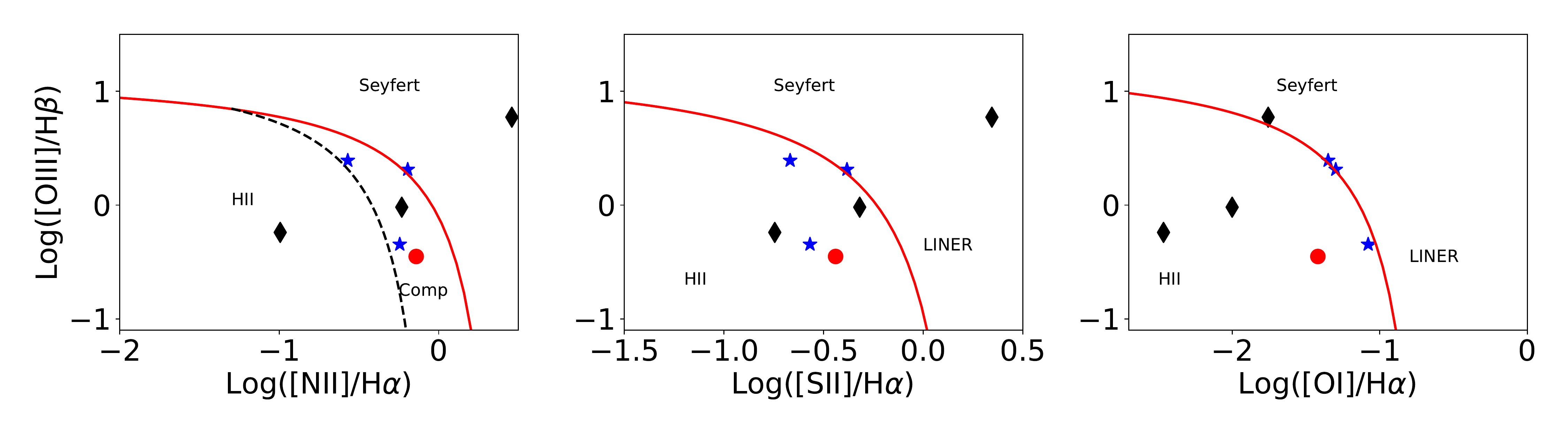}
 \caption{Diagnostic diagrams based on optical emission line ratios of [N\,\textsc{ii}] (left), [S\,\textsc{ii}] (middle) and [O\,\textsc{i}] (right). The red solid lines are used to distinguish between star-forming, Seyfert and LINER sources \citep{Kewley2001}, while the dashed line denotes composite (star-forming + AGN) systems and is taken from \citet{Kauffmann2003}. Circles represent the UV/optical TDE hosts, while stars and diamonds show the soft and hard X-ray selected TDE hosts, respectively. We show the following sources: D23--H1 (optical), NGC 5905, 2MASX J0249, SDSS J0159 (soft X-ray), UGC 1791, NGC 6021 and PGC 015259 (hard X-ray).}
\label{fig:bpt}
\end{figure*}

The emission line ratios of forbidden nebular lines, such as [N\,\textsc{ii}]\,$\lambda\lambda$\,6549, 6585, [S\,\textsc{ii}] $\lambda\lambda$\,6317, 6331 and [O\,\textsc{i}]\,$\lambda$\,6301, in combination with H$\alpha$ and H$\beta$ have proven to be reliable indicators of the source of the ionising radiation field in the nuclei of galaxies. In particular, we use the Baldwin--Phillips--Terlevich (BPT; \citealt{Baldwin1981}) diagram of the aforementioned line ratios measured from both the WHT and Keck spectra to investigate the most likely ionising field.
We note that the slit widths used probe different physical regions in the host galaxies due to their respective redshifts. We aim to identify potential AGN host galaxies using the 3 BPT diagrams (N/Si/O), although for some sources not all line ratios can be measured. We present measurements of the sources (where available) in 3 diagrams in Figure \ref{fig:bpt}. Different symbols indicate the nature of the TDE selection: circles for UV/optical (1 source), stars for soft X-ray selected (3 sources) and diamonds for the hard X-ray selected (3 sources) candidates. The host galaxies of all the other TDE candidates do not show significant emission lines, indicating that they are probably quiescent (i.e. no observable star formation or nuclear activity). 

NGC 6021 is unambiguously identified as an AGN based on its emission line content. Furthermore, the measured velocity dispersion of 187 km s$^{-1}$ translates to M$_{\rm BH}$\,=\,1.2\,$\times$\,10$^8$\,M$_{\odot}$, which is in excess of the Schwarzschild-Hills mass of a solar-type (or lower mass) star. Two other sources, namely PGC 015953 and SDSS J0159 (see also \citealt{Merloni2015} and Section \ref{sec:individualobjects}), are unambiguously identified as composite SF+AGN nuclei, while the rest is in the SF region of the diagrams. For 3XMM J1500, our GTC spectrum does not cover H$\alpha$ but \citet{Lin2017} report that it falls in the SF region. This disfavours AGN activity in the other host galaxies as the likely source for the variability in X-rays, with the notable caveat that for some sources our slit widths were wide (up to 1.5 arcsec), and nuclear star formation can potentially outshine the AGN (see e.g. \citealt{Gezari2003}). High spatial resolution spectroscopic observations are required to unambiguously determine the nuclear emission line content.  Nevertheless, 19 out of 27 X-ray TDE host galaxies presented here do not have any observable emission line content, nor persistent X-ray emission \citep{Auchettl2017, Walter2016} making AGN activity an unlikely explanation for the majority of events.

\subsection{Black hole masses}
\begin{figure*}
 \includegraphics[width=\textwidth]{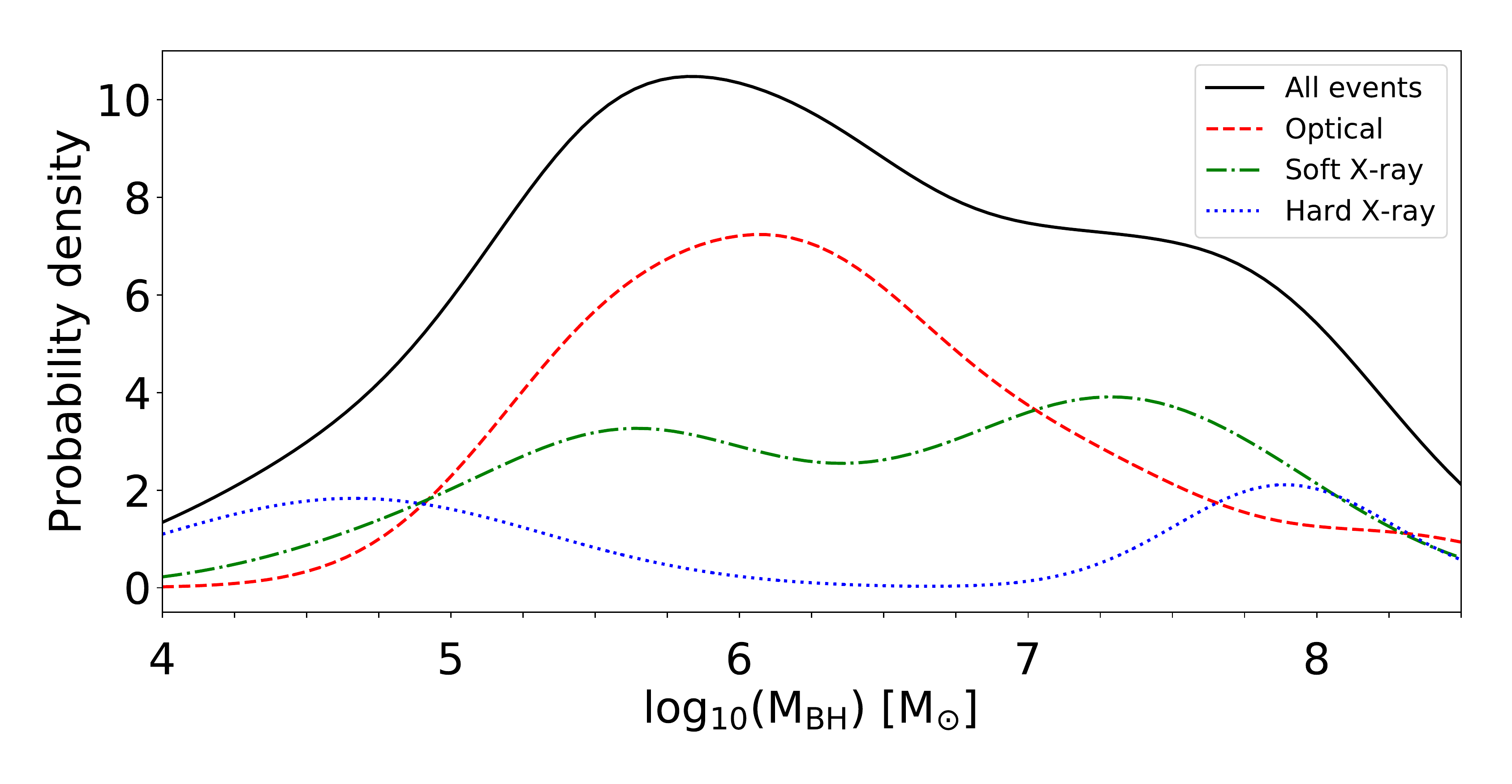}
 \caption{Mass distribution of TDE host galaxies obtained by kernel density estimation, using a kernel width equal to the uncertainty in M$_{\rm BH}$. The optical sample (red dashed line) peaks near 10$^{6}$\,M$_{\odot}$, while the X-ray distribution (green dot-dashed line) appears much flatter. However, KS and Anderson-Darling tests do not find statistically significant differences between these observed distributions; each pair of distributions is maginally consistent with being drawn from the same parent population.}
\label{fig:bhmassdistribution}
\end{figure*}

The BH mass distribution is presented in Figure \ref{fig:bhmassdistribution}. We use a kernel density estimation to take into account the uncertainties in the velocity dispersion measurements as well as the scatter in the M\,--\,$\sigma$ relation. In particular, we use a Gaussian kernel with kernel width equal to the uncertainties in M$_{\rm BH}$ quoted in Table \ref{tab:results} to represent each measurement as a probability density function (pdf). These pdfs are then summed over the relevant samples to obtain the distributions. We show the combined UV/optical (15 sources) + soft X-ray (11 sources) + hard X-ray (5 sources) mass distribution as a solid black line, while the red dashed, green dot-dashed and blue dotted lines represent the UV/optical, soft and hard X-ray selected sources, respectively. We regard ASASSN--14li and ASASSN--15oi as optical TDEs, as that is the wavelength domain they were discovered in. We reiterate that the sources at low $M_{\rm BH}$ should be interpreted with caution due to the lack of calibration of the $M_{\rm BH}-\sigma$ relation at these masses. On a side note, we confirm the discrepancy in black hole mass for PS1--10jh from velocity dispersion measurements and lightcurve modelling \citep{Mockler2018}. Using the new GTC spectra, we measure $\sigma$\,=\,60\,$\pm$\,3 km s$^{-1}$, consistent within the uncertainties with the value of 65\,$\pm$\,3 km s$^{-1}$ reported in \citet{Wevers2017}.

We perform a 2-sample Kolmogorov-Smirnov (KS) test to reject the null hypothesis that the mass distributions are drawn from the same parent distribution, and find that we cannot reject it at high significance. However, a KS test may not be the most appropriate test to use because the distributions of the different samples have similar mean values. The sample sizes are such that we do not expect significant biases in black hole mass due to Poisson statistics, and there is no evidence for systematic biases against low- or high-mass SMBHs in the X-ray sample.
From pair-wise comparisons, we find a p-value of $p$\,=\,0.43 for the optical and soft X-ray samples, $p$\,=\,0.30 for the optical and hard X-ray samples, while for the soft and hard X-ray samples we find $p$\,=\,0.33 and finally for the optical and combined soft+hard X-ray samples we find $p$\,=\,0.36. In other words, we cannot reject the null hypothesis that these distributions are drawn from the same parent distribution using the KS test.

As an alternative, we also use an Anderson-Darling test, which is more sensitive to the wings of the distribution for samples with similar values of the mean/mode. In this case, we find test statistics corresponding to p-values of $p$\,=\,0.28, 0.04, 0.06 and 0.12 (see also Table \ref{tab:statstests}) for optical--soft X-ray, optical--hard X-ray, soft--hard X-ray and optical--X-ray, implying there is no robust statistical evidence for the optical and soft X-ray samples being drawn from different parent populations. There is some marginal evidence (at the $\sim$\,2\,$\sigma$ level) that the hard X-ray sample is drawn from different parent distributions than the optical and soft X-ray samples. However, larger samples of sources are needed to robustly characterise differences in the mass distributions, in particular for the hard X-ray sample which consists of 4 sources. 

\begin{table}
\centering
\caption{P-values from an Anderson-Darling test for different host properties, with the null hypothesis that they are drawn from the same parent distribution. A p-value of 0.05 indicates that we can reject this hypothesis at the 2\,$\sigma$ level. The X-ray sample is the combination of the soft and hard X-ray sources. }
\begin{tabular}{cccc}
\hline
Sample & M$_{\rm BH}$ & Stellar mass & Abs. mag \\\hline
Optical -- soft X-ray & 0.28 & 0.19 & 0.33 \\
Optical -- hard X-ray & 0.04 & 0.04 & 0.04\\
Soft X-ray -- hard X-ray & 0.06 & 0.17& 0.22\\
Optical -- X-ray & 0.12 &0.04& 0.08 \\\hline
\end{tabular}
\label{tab:statstests}
\end{table}

\subsection{Galaxy stellar masses and bulge masses}
\label{sec:comparison}
We use SDSS photometry (where available) to calculate the total stellar mass and absolute g-band magnitude of the host galaxies, using the \textsc{kcorrect} software \citep{Blanton2007}. If no SDSS data is available, we use Pan-starrs (PS1) data instead. We assume H$_0$\,=\,0.7 when fitting the photometry, and have corrected for Galactic dust extinction using the \citet{Schlegel1998} dust maps. The results are presented in Table \ref{tab:results}. The typical stellar mass content of the host galaxies, irrespective of the TDE selection criterion, is 10$^{9.5}$\,--\,10$^{10.5}$ M$_{\odot}$ (see also table 1 in \citealt{vanvelzen2018}). There are, however, several (both hard and soft) X-ray TDE hosts that are significantly less luminous and/or less massive (in terms of stellar mass) than the least massive and faintest optical TDE host (PS1--10jh, which has M$_{\rm g}$\,=\,--18.1 and log(M$_*)$\,=\,9.5 M$_{\odot}$). As for the black hole masses, we perform KS and Anderson-Darling tests for the distributions of stellar mass and host absolute magnitude in the g-band, to reject the hypothesis that the different samples have properties drawn from the same parent distribution. The results are summarised in Table \ref{tab:statstests}; there is suggestive evidence that the hosts of the hard X-ray sample differ in all three properties from the optical and soft X-ray samples.

In particular there are three hard X-ray selected TDE candidate hosts, UGC 1791, PGC 1127938 and PGC 1185375, that have absolute magnitudes around M$_{\rm g}$\,=\,--16 or fainter and are morphologically very similar to each other, lacking a clear bulge component. The latter two sources show no emission lines, whereas UGC 1791 is classified as a SF galaxy. Because the M\,--\,$\sigma$ relation is increasingly uncertain at the low velocity dispersions measured for these galaxies, we provide an alternative estimate using the relation between total stellar mass and black hole mass \citep{Reines2015}. This results in M$_{\rm BH}$ in the range $\sim$\,10$^{5.5 \pm 0.47}$ M$_{\odot}$ for these galaxies. To provide a more quantitative estimate of the central concentration, we use the bulge-to-total (B/T) g-band flux ratios (assuming a classical bulge) from \citet{Simard2011} for PGC1185375 (0.15) and PGC1127938 (0.40). For UGC1791, we estimate the (B/T) ratio in the g-band by taking the ratio of the PS1 PSF and Kron fluxes to estimate (B/T)$_g$\,=\,0.15. Although a rigorous comparison is not possible with these estimates, it seems that while PGC1127938 is consistent with a higher than usual B/T ratio seen in other TDE hosts, the other two sources have more typical low B/T values as seen in SDSS galaxies at similar BH masses (figure 5 in \citealt{Law-Smith2017}).

\begin{figure}
 \includegraphics[width=0.51\textwidth]{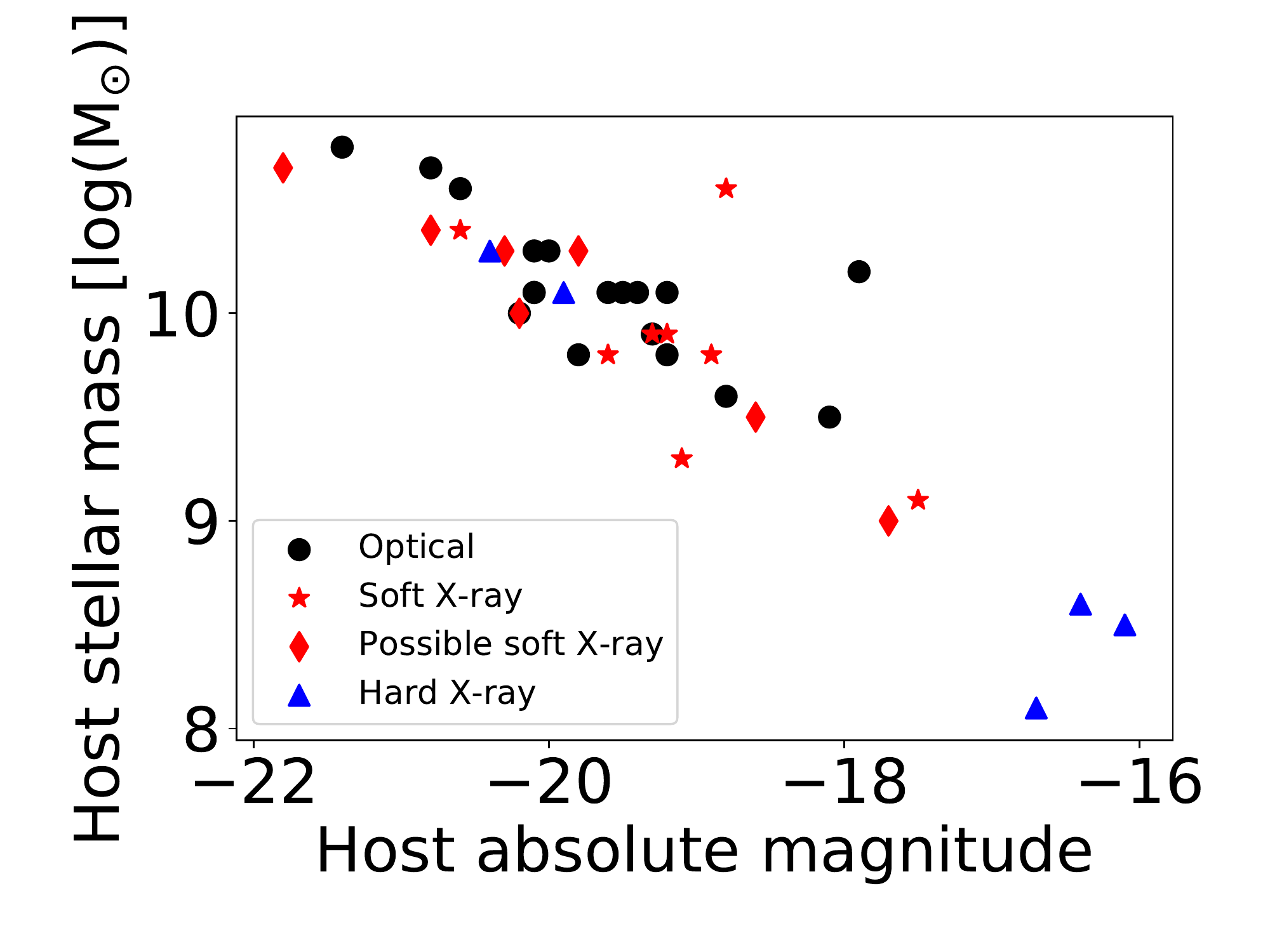}
 \caption{Host galaxy stellar mass as a function of galaxy absolute magnitude for the respective samples. Several hard X-ray events are stark outliers in this parameters space compared to the UV/optical events, while some soft X-ray hosts appear to fill the gap between the two groups. }
\label{fig:scatterplot}
\end{figure}
The hosts of two soft X-ray selected events, 2MASX J0249 and RBS 1032, are less extreme outliers in terms of galaxy absolute magnitude and stellar mass when compared to the optical TDE hosts, although they are still fainter and less massive than the host of PS1--10jh (Figure \ref{fig:scatterplot}). These 2 galaxies are significantly more centrally concentrated; we estimate that for RBS1032, (B/T)$_g$\,=\,0.38\footnote{A (B/T)$_g$\,=\,0.75 is reported in \citealt{Law-Smith2017}.} and for 2MASX J0249, (B/T)$_g$\,=\,0.39. Although it has been predicted that X-ray TDEs should preferentially occur around less massive SMBHs (which generally inhabit less massive and fainter galaxies; \citealt{Dai2015}), we find no clear systematic differences for the populations of soft X-ray and UV/optical selected TDEs as a whole in terms of galaxy luminosity or stellar mass (Table \ref{tab:results}) nor black hole mass. 
Instead, assuming that all events belong to the same parent population, there is a continuum in host galaxy properties, and our data does not show a clear systematic trend for X-ray and optical TDEs to occur in distinct host galaxy populations. This is in agreement with our current understanding of these events \citep{Law-Smith2017, Graur2018}.

\begin{figure}
 \includegraphics[width=0.51\textwidth]{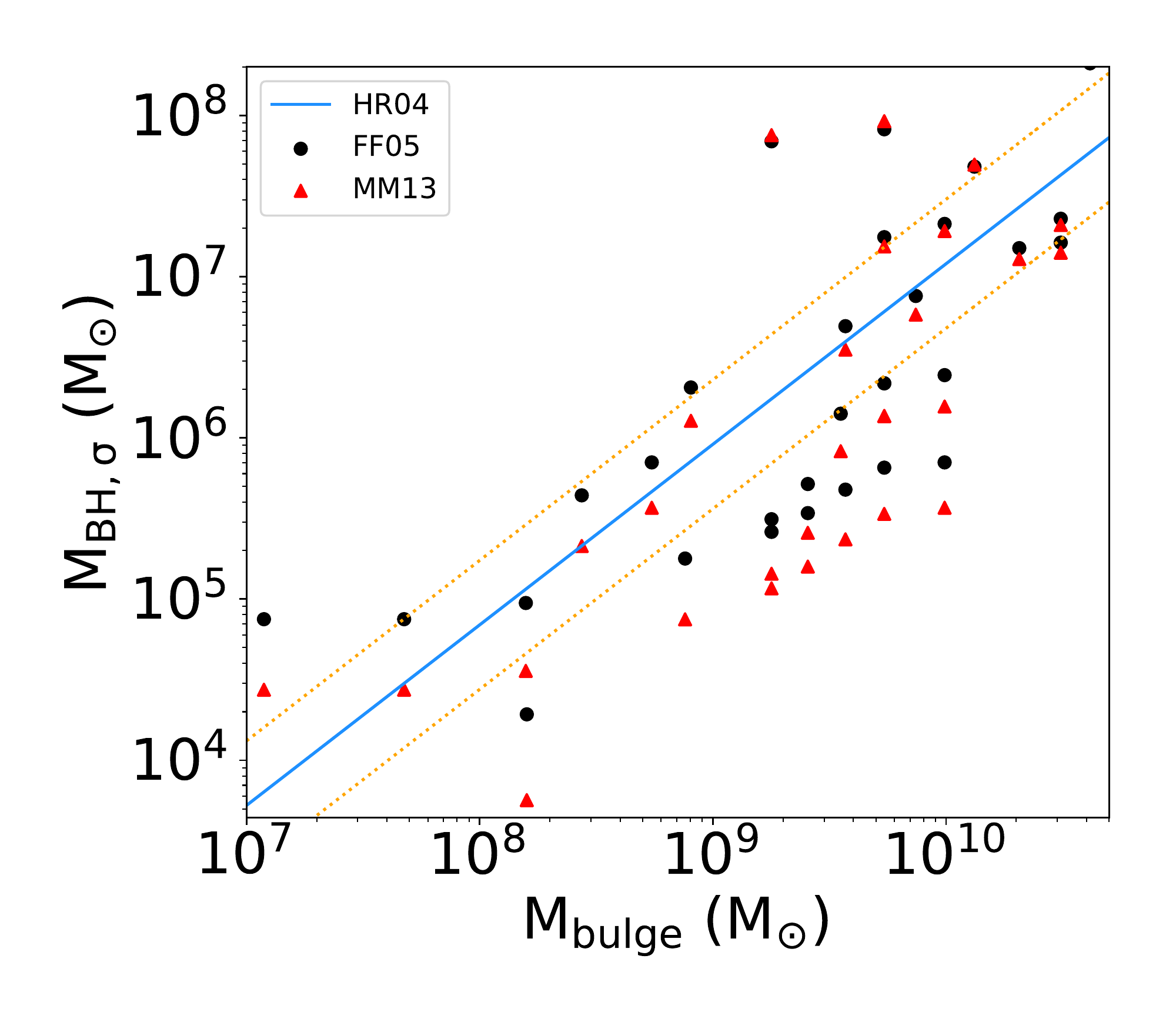}
 \caption{Galaxy bulge mass as a function of black hole mass, derived using two different M\,--\,$\sigma$ relations (\citet{Ferrarese2005} in black circles, and \citet{Mcconnel2013} as red triangles). The solid line indicates the canonical relation between M$_{\rm BH}$ and M$_{\rm bulge}$ by \citet{Haring2004}. The dashed lines denote the typical scatter of 0.4 dex for this M$_{\rm BH}$\,--\,M$_{\rm bulge}$ relation. The three left-most sources are the hard X-ray sources with low measured velocity dispersions.}
\label{fig:msigmambulge}
\end{figure}

We can use the total stellar mass content to estimate the galaxy bulge mass, which is known to correlate with the central black hole mass (e.g. \citet{Haring2004, Sani2011}). To this end, we estimate the bulge mass of each galaxy by correcting the total galaxy stellar mass with a measured \citep{Simard2011} or an empirically derived B/T ratio (\citealt{Stone2018}, their table B1). We note that the latter values are averages per stellar mass bin, and the actual values have a large scatter around this mean. We linearly interpolate between the tabulated values to estimate the appropriate correction factor. The results are shown in Figure \ref{fig:msigmambulge}. The solid line represents the relation found by \citet{Haring2004}, and the dashed lines represent the typical 0.4 dex scatter. \citet{Wevers2017} noted that the black hole masses for TDE hosts obtained using the M\,--\,$\sigma$ relation were downward revisions of literature masses, which were mostly based on bulge luminosities. Here we find that the measurements using galaxy bulge mass yield black hole masses that are largely in agreement with the masses derived from velocity dispersions for the range M$_{\rm BH}$\,=\,10$^5$\,--\,10$^{8}$. Some sources are found at higher scatter (consistent with the scatter of the \citet{Haring2004} relation), but no systematic trends are evident.

\subsection{Notes on individual objects}
\label{sec:individualobjects}
\subsubsection{2MASX J0249}
\citet{Esquej2007} discovered this TDE candidate in the XMM-Newton slew survey, and tentatively identified a broad base to the H$\alpha$ line in a WHT/ISIS spectrum, taken on 2006 August 18. The SNR of the spectrum is low, and the identification was therefore ambiguous. Our Keck spectrum does not show any signs for a broad component to H$\alpha$, and in the BPT diagrams it is firmly in the SF region. We reanalyse the data presented by \citet{Esquej2007}, obtained from the ING archive\footnote{http://casu.ast.cam.ac.uk/casuadc/ingarch/query}, and find that although the broad component is also visibly present in our reanalysis, given the low signal-to-noise ratio (SNR) there is no statistical evidence of this component being real rather than noise. Given the significant amount of time (2 years) between the flare detection in soft X-rays and the optical spectrum, it seems unlikely that this would be the optical spectroscopic signature of the TDE. On the other hand, this example does show how TDEs might masquerade in single-epoch spectroscopic surveys as ambiguous classifications of host galaxies with narrow emission lines, leading to a potential selection bias against finding such events in large spectroscopic surveys.

\subsubsection{SDSS J0159}
SDSS J0159 was identified by \citet{Lamassa2015} as a changing-look AGN, changing appearance in its optical spectrum with the broad Balmer lines disappearing as the X-ray luminosity and AGN continuum flux decreased by a factor of 6. These authors derive a black hole mass based on the FWHM of the broad component of the H$\beta$ line of 2\,$\times$\,10$^8$\,M$_{\odot}$. On the other hand, \citet{Merloni2015} argue that this event is consistent with a TDE based on the lightcurve evolution and energetics.
Here we determine the velocity dispersion based on absorption features, which implies a black hole mass of $\sim$\,10$^{7.2}$\,M$_{\odot}$, significantly lower than the broad line estimate of \citet{Lamassa2015}. This could suggest that the broad transient emission lines were the optical signature of a TDE, in which case no clear correlation between M$_{\rm BH}$ and the emission line FWHM would be expected. The SDSS spectrum showing the broad H$\alpha$ line was indeed taken near the peak of the optical lightcurve \citep{Merloni2015}.
Fitting the narrow emission lines with Gaussian profiles yields a similar velocity dispersion measurement. However, with the relatively high resolution of our Keck spectrum, the narrow emission lines are clearly resolved into two narrower components. All emission lines are masked during the absorption line fitting, such that this does not affect the measured velocity dispersion. For the H$\alpha$ emission line, we measure a peak-to-peak separation of $\sim$\,150 km s$^{-1}$; all emission lines show evidence for two kinematically distinct components. These two components could originate from the rotation of the narrow line region, and were unresolved in all the spectra presented in \citet{Lamassa2015}. Other explanations include galactic scale outflows (e.g. \citealt{Greene2011}) or merging galaxy pairs (where the 2 components originate from the NLR of each SMBH in a dual AGN system; e.g. \citealt{Comerford2009}).

\section{Discussion}
\label{sec:discussion}
\subsection{A bias against AGN host galaxies?}
The absence of TDEs in AGN host galaxies could be the result of a selection bias in the spectroscopic follow-up, but could potentially also be related to the dust content of AGNs (as compared to quiescent galaxies). The nuclear black holes in type 2 (narrow-line) AGN are by definition surrounded by a thick dusty structure obscuring the inner (broad line) region and SMBH, which would inhibit the detection of nuclear flares at UV/optical wavelengths. For type 1 AGN, intrinsic optical variability and the presence of persistent broad lines could inhibit the detection of the typical TDE signatures such as a fast rise exponential decay lightcurve and transient broad ($\sim$\,10$^4$ km s$^{-1}$) H and He emission lines. The bias in the X-ray sample could also be partially ascribed to misclassification of variability in known AGNs. Nevertheless, given the close connection between galaxy mergers, the fuelling of the central black hole in AGN, and star formation, it is expected that the TDE rate in these active galaxies should be comparable to or higher than in quiescent galaxies (e.g. \citealt{Karas2007, Kennedy2016}). Although AGN are relatively rare among the local galaxy population (accounting for about 5\,--\,10\,\% of the total number of galaxies, \citealt{Kauffmann2003}), mQBS galaxies are comparatively even rarer ($\sim$\,2\,\% of the galaxy population). The BPT diagrams indicate that we are currently largely missing these events. 

The observed over-representation of TDEs in rare E+A galaxies may connect to TDE rates in AGN, as the E+A evolutionary state is often reached following a period of intense star formation after a merger \citep{Zabludoff1996}. The post-merger evolution likely includes an AGN phase that expels the remaining gas and thereby quenches the star burst \citep{Hopkins2006}, after which unusual stellar dynamical processes can enhance the TDE rate. It is not yet clear which of the many proposed dynamical processes \citep{Arcavi2014, Stone2016, Stone2018, Madigan2018} is predominantly responsible for elevating post-starburst TDE rates, but many of them are expected to arise prior to the cessation of star formation. Post-merger galaxies may therefore display similarly elevated TDE rates at all phases, including in the immediate progenitors of E+A galaxies, which include star-forming (e.g. \citealt{Tadhunter2017}) and AGN hosts. 

If the bias against TDEs in AGN is purely observational, a systematic and unbiased survey for TDE signatures in AGN host galaxies could help improve our understanding of the dynamical (or other) mechanisms that are responsible for the observed elevated TDE rate in E+A galaxies. Quantifying the TDE rate in terms of nuclear dust obscuration (e.g. E(B--V) of the host galaxies) could show whether large scale gas/dust columns play a role in TDE observability.

\subsection{Predictions for and correlations with black hole mass}
\label{sec:predictions}
Although accurate predictions for the black hole mass distribution in the literature are scarce, some models do predict their preferred M$_{\rm BH}$ distribution and/or correlations of other observables such as the temperature and peak luminosity with black hole mass. We briefly explore these predictions here, starting with the black hole mass distribution.

\citet{Dai2015} argue that thermal soft X-ray emission will only be produced around (i) low-mass SMBHs (M$_{\rm BH} \lesssim 10^{6.7}$\,$M_{\odot}$), where disk effective temperatures can reach into the X-rays, and (ii) in the case of a relativistic pericenter, which, for small SMBHs, implies a deeply plunging orbit. Although we can not constrain the depth of disruption, the mass distribution of the soft X-ray sample, which is roughly flat in M$_{\rm BH}$ up to $10^8$\,$M_{\odot}$, seems incompatible with this prediction, perhaps indicating the presence of higher temperature disks due to large Kerr spin parameters and/or spectral hardening corrections. Alternatively, contamination of the X-ray sample by AGN could bias the distribution towards higher masses (but see Section \ref{sec:bpt}).
 
At higher SMBH masses, all TDE pericenters are sufficiently relativistic to enable rapid circularization, unless stream crossings are impeded by Lense-Thirring precession \citep{Guillochon2015, Hayasaki2016}. 
If high inner disk accretion rates are required to power UV/optical light curves (e.g. in reprocessing models such as \citealt{Dai2018}), one would then expect some bias towards higher mass SMBH hosts for UV/optical flares, although this bias is modest (e.g. figure 11 in \citealt{Stone2016}).
This also appears at odds with the current observations of both the UV/optical and X-ray sample (Figure \ref{fig:bhmassdistribution}). 

\begin{figure}
 \includegraphics[width=0.5\textwidth]{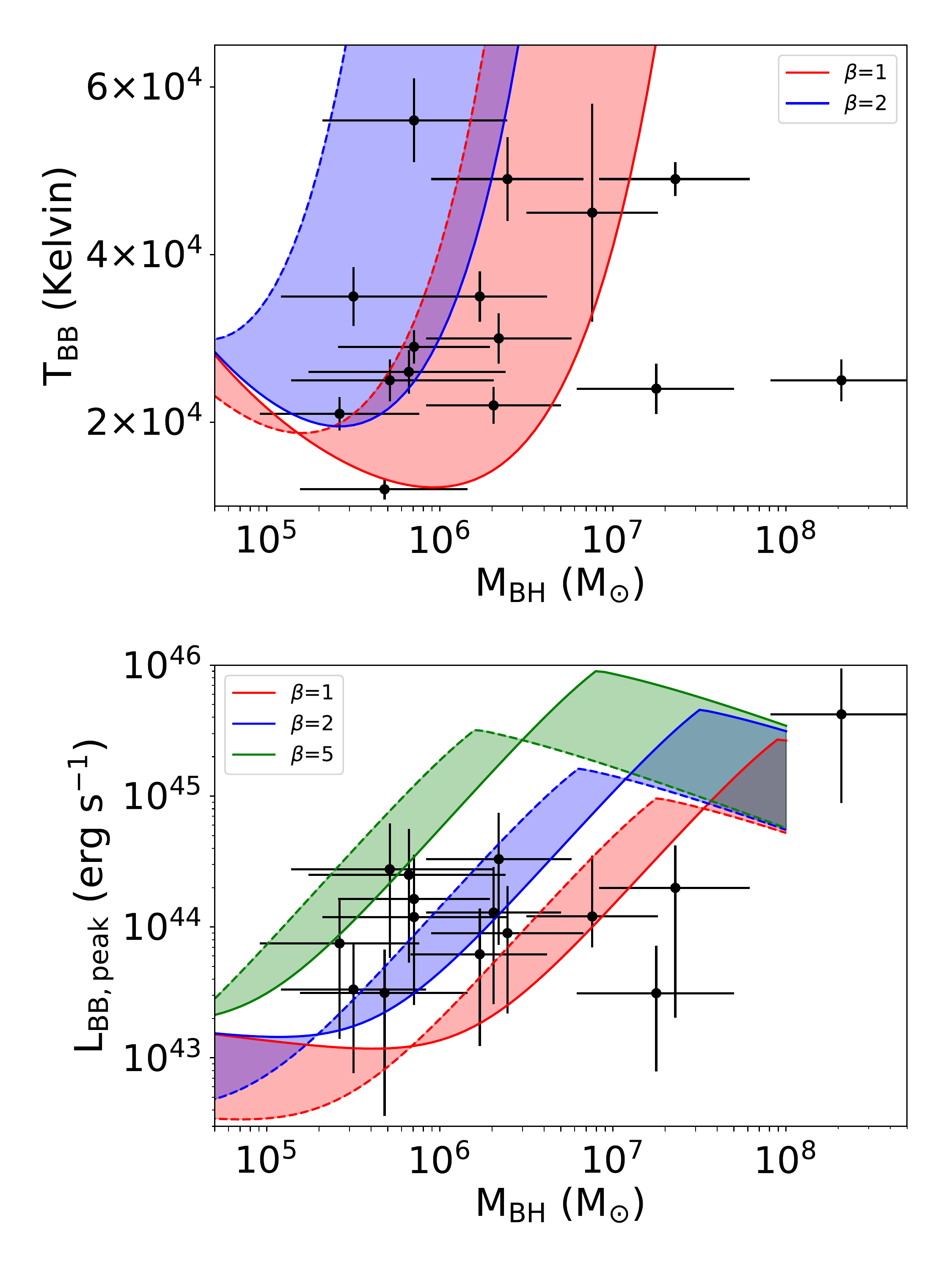}
 \caption{Optical blackbody temperature and luminosity at peak as a function of black hole mass for the UV/optical events. No clear correlations are observed. We overplot predictions from the shock model for different impact parameters and stellar masses (the solid lines represent a 1\,M$_{\odot}$ star, the dashed lines a 0.1\,M$_{\odot}$ dwarf.}
\label{fig:correlations}
\end{figure}

In terms of correlations between observables, if the UV/optical emission is produced in stream-stream collisions, \citet{Piran2015} predict that the temperature of the emission should scale inversely with M$_{\rm BH}$ when streams self-intersect near apocenter, although a large spread in blackbody temperature will exist in a sample with varied penetration factors. We show the peak blackbody temperature as a function of black hole mass in the top panel of Figure \ref{fig:correlations}, from which we conclude that no such inverse scaling is observed. The bottom panel of Figure \ref{fig:correlations} shows the peak luminosity in the g-band as a function of black hole mass. 

We overplot basic predictions of this model, assuming that a spherical photosphere at the self-intersection radius (R$_{\rm SI}$) radiates a luminosity
\begin{equation}
\rm L = \frac{\dot{M}_{dyn}\ G\ M_{BH}}{R_{SI}}
\end{equation}
where $\dot{\rm M}_{\rm{dyn}}$ is the dynamical mass fall-back rate at peak and G the gravitational constant. This provides an upper limit for the shock-powered luminosity, as it assumes that all the stream kinetic energy is thermalised and radiated. This luminosity will be lower if i) orbital plane precession from misaligned SMBH spin makes the stream self-intersections grazing \citep{Guillochon2015, Hayasaki2016}, ii) some of the energy is converted into kinetic energy of material piling up at R$_{\rm SI}$, iii) there is a strong surface density mismatch between the inbound and the outbound stream when they self-intersect. 

The measurements in both temperature and luminosity are broadly consistent with low penetration factors, $\beta$\,$\sim$\,1\,--\,2. As just explained, the luminosity from shocks in our simple model is likely a conservative upper limit, and depending on the details of the post-disruption dynamics may be significantly lower. In this respect, we note that if the luminosity is lower by as little as a factor of $\sim$\,2, several events require $\beta$\,$>$\,2 in order to remain consistent with the shock-powered model. More detailed model predictions are required to investigate quantitative (dis)agreements with observations.

Finally, we will see in Section \ref{sec:emissionradius} that the shock model also correctly predicts the UV/optical emission radius.

\subsection{Eddington ratio of the UV/optical and X-ray emission}
\begin{figure}
 \includegraphics[width=0.52\textwidth]{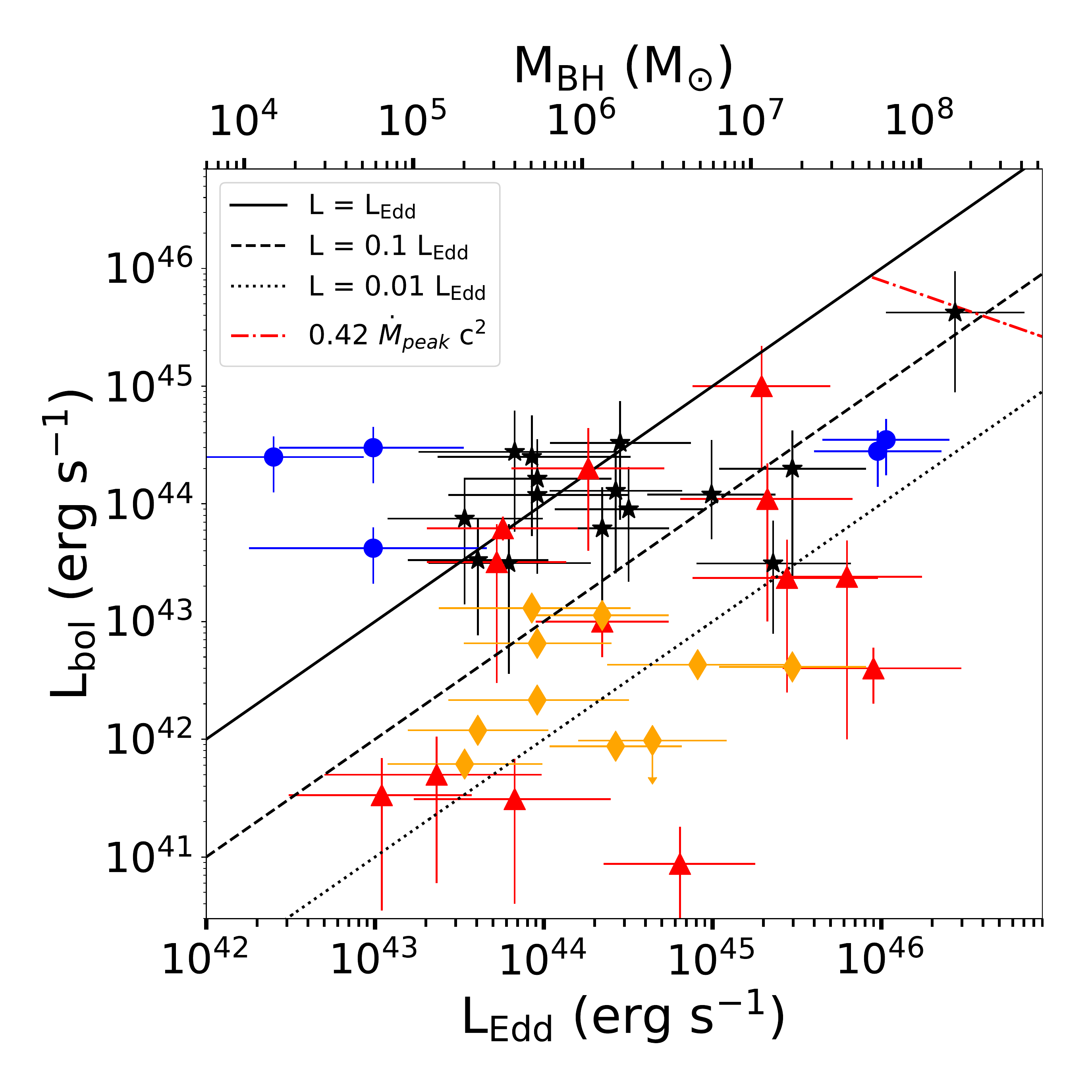}
 \caption{Eddington ratio of the early time UV/optical and X-ray emission observed in TDEs at peak. Black stars indicate the UV/optical emission; red triangles soft X-ray emission, blue circles hard X-ray emission and orange diamonds UV late-time emission. No bolometric correction is applied for the X-ray measurements. The upward sloping lines indicate different Eddington ratios to guide the eye, while the dot-dashed red line indicates the peak fall-back luminosity for a maximally spinning Kerr black hole.}
\label{fig:eddingtonratio}
\end{figure}

We compute the integrated blackbody UV/optical emission using the blackbody temperature and peak absolute magnitude in the g-band as in \citet{Wevers2017}. The X-ray measurements are taken from \citet{Auchettl2017} and \citet{Walter2016} for the soft and hard X-ray samples, respectively. No bolometric corrections are applied to the X-ray measurements.
In Figure \ref{fig:eddingtonratio} we plot the optical measurements as black stars, while the red triangles and blue circles represent the soft and hard X-ray selected events, respectively. The late-time UV measurements from \citet{vanvelzen2018b} are shown as orange diamonds. While the optical events tend to have Eddington ratios in excess of $\sim$\,0.2\,L$_{\rm Edd}$, the X-ray events appear to cluster at lower Eddington ratios, ranging from 10$^{-4}$ to 0.05 L$_{\rm Edd}$. The average Eddington ratios of the optical and soft X-ray selected events are 1\,$\pm$\,1 and 0.27\,$\pm$\,0.4, respectively (this does not include the two super-Eddington events Swift J1644 and Swift J2058, which are not considered in this work). For completeness we note that although the implied Eddington ratios for the three hard X-ray TDE candidates in dwarf galaxies are greater than 1, if we assume the black hole mass estimates from the \citet{Reines2015} relation ($\sim$\,10$^{5.5}$\,M$_{\odot}$) to be more representative, their Eddington ratios are consistent with unity. It should be kept in mind that the X-ray lightcurves (and indeed, some of the lightcurves of the optical sample) are typically poorly sampled, and thus these estimates represent lower limits on the true peak luminosity.

Several soft X-ray selected events clearly stand out from the rest of the soft X-ray sample near the Eddington limit of their host black holes: SDSS J1201, SDSS J1323, 3XMM J1500 and 3XMM J1521. SDSS J1201 and SDSS J1323 were identified in \citet{Auchettl2017} to have typical energy release times similar to the jetted events Swift J1644 and Swift J2058, while other aspects of the emission (such as the hardness ratio and power-law index) are similar to ASASSN--14li, which likely launched a mildly relativistic jet \citep{Pasham2018}. While the Swift events had a highly super-Eddington plateau phase of X-ray emission, the fact that J1201 and J1323 are near the Eddington limit implies that the emission was not highly relativistic (hence the X-ray emission is thermal in nature), consistent with the findings of \citet{Auchettl2017} that the hardness ratios support thermal disk emission for these events. 3XMM J1500 was also identified by \citet{Lin2017} as a likely TDE with a 10 year super-Eddington phase.

The peak UV/optical (blackbody) emission of TDEs is consistent with being Eddington limited (Figure \ref{fig:eddingtonratio}), and furthermore consistent with being produced in a region similar to the stream self-intersection radius, several 100s of gravitational radii from the SMBH. Several suggestions as to the nature of the optical/UV emission exist, including the reprocessing of accretion power in a quasi-static debris layer \citep{Loeb1997, Guillochon2014, Coughlin2014, Roth2016}, reprocessing in an outflow \citep{Strubbe2009, Lodato2011, Miller2015, Metzger2016, Roth2018}, and shock-powered emission from self-intersection debris streams \citep{Piran2015, Shiokawa2015}. In the first two frameworks, emission should naturally be capped near the Eddington limit, as is observed; such a limit does not exist {\it a priori} for shock-powered emission, but in practice, predicted luminosities in the shock-powered model are almost always sub-Eddington compared to the SMBH Eddington limit \citep{Piran2015}. We note, however, that the emission region of the UV/optical radiation is located 10\,--\,100 ISCO radii from the SMBH (Figure \ref{fig:bbradii}), and the {\it local} Eddington limit for radiation produced in-situ at the self-intersection point is expected to be 10$^2$\,--\,10$^4$ times lower than the limit in the vicinity of the SMBH. This effectively makes the observed UV/optical emission highly super-Eddington if produced locally at the self-intersection point. The fact that the UV/optical emission appears to be capped at the SMBH Eddington limit seems to be artificial in the shock-powered scenario, which suggests that this is not the powering mechanism of this radiation component. The Eddington limited UV/optical emission instead suggests that the radiation is related to the SMBH.

ASASSN--15lh, a source whose nature is still debated \citep{Dong2016, Leloudas2016, Margutti2017}, is not an outlier when compared to other optical TDEs in terms of its Eddington ratio. If we use a simple dynamical prediction for the peak fall-back rate (e.g. \citealt{Stone2013}), and furthermore assume the maximum radiative efficiency for a spinning black hole (i.e. $\eta$\,=\,0.42), we find that the emission of ASASSN--15lh is indeed consistent with the predicted Eddington ratio and luminosity (the red dot-dashed line in Figure \ref{fig:eddingtonratio}). 

Using the peak X-ray luminosity measurements from \citet{Auchettl2017}, we can now calculate the Eddington ratios of the soft and hard X-ray selected TDE candidates at the observed peak of the lightcurve and compare the results. We also include late-time UV measurements of a sample of 10 UV/optical discovered events presented in \citet{vanvelzen2018b}. From the latter measurements, these authors conclude that viscously spreading accretion disks are present at late times (5\,--\,10 years after peak), with the emission inconsistent with a simple power-law decay extrapolated from the early lightcurve. This inconsistency with a single power-law decay is also observed in the X-ray lightcurves \citep{Auchettl2017}. We apply a (model-independent) bolometric correction factors based on the ratio of the UV luminosity \citet{vanvelzen2018b} and the integrated UV/optical luminosity. In Figure \ref{fig:eddingtonratio} these are shown as orange diamonds; their Eddington ratios are similar to those of the X-ray observations, although for the latter no bolometric correction is applied.

If returning tidal debris i) circularises promptly and ii) forms an unobscured accretion flow, then simple disk models predict thermal soft X-ray emission at Eddington or higher levels \citep{Ulmer1999, Lodato2011}. The relatively low observed X-ray Eddington ratios, L$_{\rm X} \sim 0.01$ L$_{\rm Edd}$, suggest that one of these two assumptions is incorrect (it is unlikely that the true peak X-ray luminosities are one or two orders of magnitude higher due to the sparse temporal sampling, given the observed decay rates \citep{Auchettl2017}). 

TDE disks assemble rapidly when streams have relativistic pericenters and are confined to a single orbital plane \citep{Hayasaki2013, Bonnerot2016}, but disk assembly can be delayed for non-relativistic pericenters or around spinning SMBHs (see Section \ref{sec:predictions}). A more slowly assembled disk will see slower time evolution in its X-ray light curve, as is observed in most soft X-ray TDEs \citep{Auchettl2017}.

On the other hand, \citet{Mockler2018} showed that the optical events likely have short circularisation and viscous timescales, indicating that material falls back to the SMBH at super-Eddington rates after disruption. The question that needs to be addressed is, then, why do we not observe luminous X-ray radiation at early times. \citet{Dai2018} recently proposed a unified model for TDEs, where X-ray radiation is only visible when the observer is looking down the funnel of a jet or outflow (see also \citealt{Metzger2016}. The relatively large spread in Eddington ratio (10$^{-4}$ to 1) would then be the result of varying amounts of reprocessing and extinction due to variations in the covering fractions of optically thick material. This is consistent with the late-time X-ray detections of several optical TDEs (PTF--09axc, PS1--10jh, D3--13; \citealt{Auchettl2017}), when the obscuring material has had time to disperse and become optically thin to the X-ray radiation. Combined X-ray and optical observations of a large sample of TDEs are necessary to test the (early-time lightcurve) predictions of this model by quantifying the relative fractions of X-ray bright optically dim, optically bright X-ray dim and X-ray bright optically bright events, respectively. 

We now turn our attention to the physical regions in which the various emission components originate.

\subsection{The emission region of the UV/optical and X-ray emission}
\label{sec:emissionradius}
\begin{figure*}
 \includegraphics[width=0.85\textwidth]{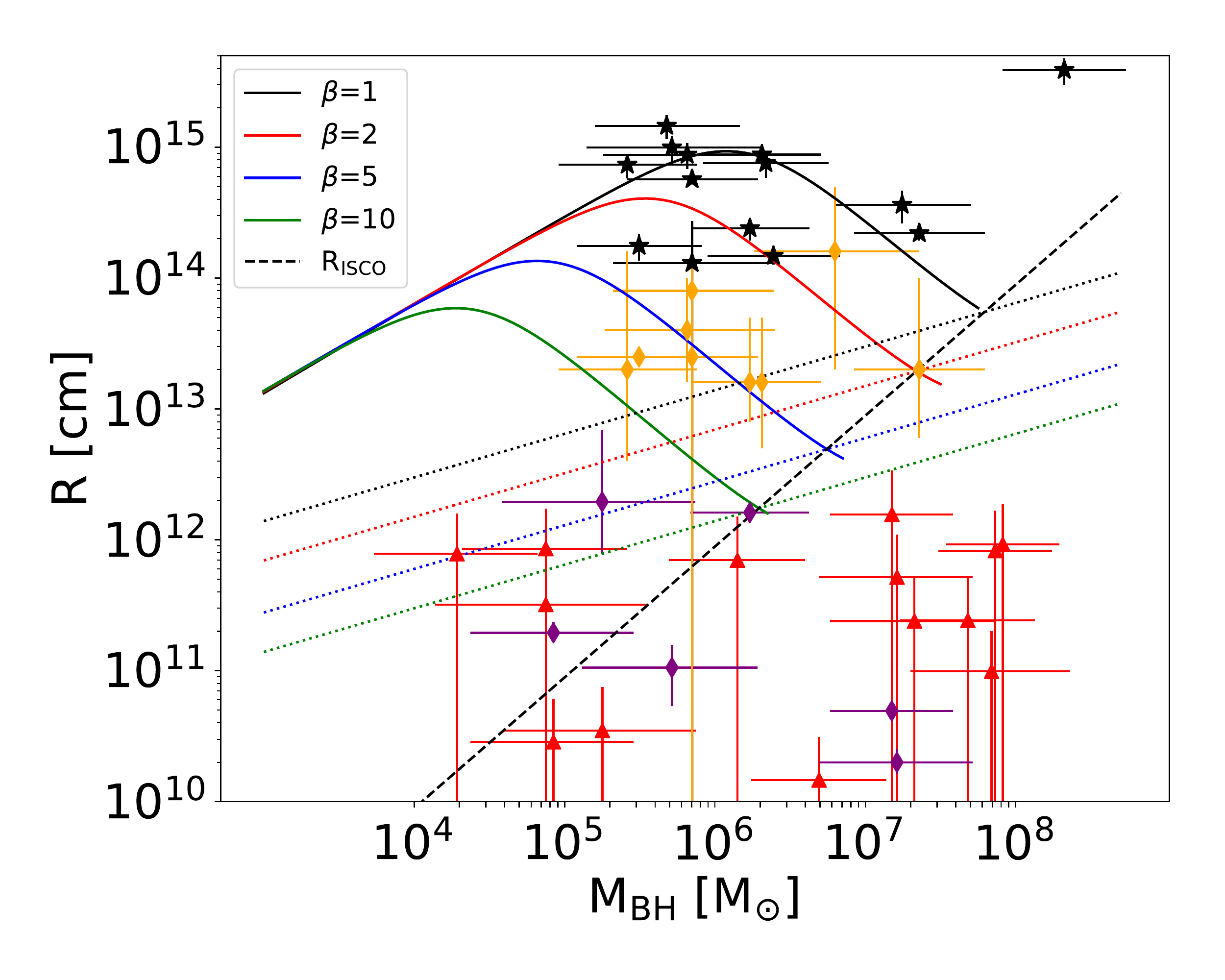}
 \caption{Emission region radius estimates for the optical (black stars), X-ray (from the peak observed luminosity as red triangles ,and from refitting the X-ray spectra as purple diamonds) and late-time UV (orange diamonds) radiation. The solid lines indicate the stream self-intersection radii at different penetration depths $\beta$, while the dotted lines indicate a compact accretion disk at 2\,$\times$\,R$_{\rm p}$. The black dashed lines shows the innermost stable circular orbit for a non-rotating BH, i.e. at 3 Schwarzschild radii. Values derived from the X-ray measurements are generally model-dependent, and in particular values below the ISCO should not be over interpreted given the data quality. }
\label{fig:bbradii}
\end{figure*}
Under the assumption of isotropic emission, \citet{Wevers2017} derive the blackbody radii of the emission regions responsible for the UV/optical early-time radiation, and compare them to some simple theoretical models, including the stream self-intersection radius and a compact accretion disk at R\,$\sim$\,2\,R$_{\rm p}$ \citep{Phinney1989}, with R$_{\rm p}$ the pericenter radius of the orbit of the disrupted star. We update their figure 9 by including the rest of the optical sample considered in this work (Figure \ref{fig:bbradii}). All the optical events cluster near regions of impact parameter $\beta$\,$\sim$\,1\,--\,2 consistent with previous results, and in continued agreement with predictions for shock-powered optical emission \citep{Piran2015}. 

Having studied soft X-ray TDE host galaxies in this work, we are now able to include the expected emission region of the soft X-ray component as well. We provide two different estimates, as follows. Our first approach is to take the peak luminosities from \citet{Auchettl2017}, and assume a typical blackbody temperature of 75 eV (similar to ASASSN--14li, \citealt{vanvelzen2016a} and ASASSN--15oi, \citealt{Holoien15oi}). We then assume that the X-rays are isotropic blackbody emission. This is unlikely to be correct for all X-ray sources, given that \citet{Auchettl2017} find several whose X-ray spectrum is better described by a power-law, and, moreover, there are no meaningful constraints on the spectral shape of the hard X-ray selected events. Nevertheless, given that the constraints on the X-ray spectral shape are not generally very strong, taking this approach allows us to make a simple approximation of the likely X-ray emission region. We assume uncertainties of 50\,\% on the peak luminosity of the hard X-ray events because of the sparse temporal sampling. As shown in Figure \ref{fig:bbradii} (red triangles), the typical emission radius we find is of order 10$^{12}$ cm or lower (Table \ref{tab:xraymeasurements}), 2\,--\,3 orders of magnitude smaller than the optical emission region. 

As a more self-consistent approach, we refit the X-ray spectra from \citet{Auchettl2017} with a 2-parameter blackbody model (we use {\sc tbabs\,$\times$\,zashift\,$\times$\,bbodyrad} as our model in XSPEC), regardless of whether this provides the best overall fit to the data. This allows us to derive the blackbody temperature and radius for each source individually. Good quality spectra are available for ASASSN--14li, ASASSN--15oi, 2MASX J0249, SDSS J1201, SDSS J0159, and RBS 1032 (Table \ref{tab:xraymeasurements}). We show these measurements as purple diamonds in Figure \ref{fig:bbradii}, and they occupy a similar parameter space as our other approximation for X-ray source size. 
In all cases, the X-ray emission regions are much more compact than the early-time optical emission regions. Our estimate of compact X-ray source size is consistent with individual object studies \citep{vanvelzen2016a, Holoien15oi}.

Interestingly, the X-ray emission region estimates are also at least an order of magnitude smaller than the radii of the viscously spreading accretion disks inferred from late-time UV observations, {\it and in many cases are smaller than plausible ISCO radii}. An effective X-ray emitting area less than that of disk annuli near the ISCO can only be explained by a large degree of obscuration, which would lend support to the reprocessing paradigm.  However, we must emphasize that these results are only suggestive, as there are many uncertainties in our X-ray source size calculations. In particular, the high reduced $\chi^2$ values suggest that a blackbody model does not describe the data very well. More detailed modeling as well as higher quality data will be needed to more accurately test whether the X-ray emitting areas are less than the effective ISCO area in TDEs.

We also overplot the measurements by \citet{vanvelzen2018b} as orange diamonds; these disks have typical radii between 10$^{13}$ and 10$^{14}$ cm, significantly smaller than the early-time UV/optical emission but significantly larger than the inferred X-ray emission region. If we assume that the X-ray emission also has its origin as accretion disk emission, it is not surprising to find the X-ray emission region being more compact than the UV emission, given that the X-rays are likely produced in the hot inner part of the disk, while the UV is produced at its outermost annuli \citep{Lodato2011}. 
%%%%%%%%%%%%%%%%%%%%%%%%%%%%%%%%%%%%%%%%%%%%%%%%%%%%%%%%%%%%%%%%%%%%%%%%%%%%%%%%%%%%%%%%%%%%%%%%%%%%%%%%%%%%%%%%%%%%%%%%%%%%%%%%%%%%%%%%%%%%%%%%%%%%%%%%%%%
\section{Conclusions}
\label{sec:conclusions}
We have analysed new and archival spectroscopic observations of 21 X-ray TDE host galaxies, as well as 17 UV/optical TDE hosts (we present new observations for 6 of these latter sources). We find that a majority of X-ray TDEs occurred in quiescent host galaxies, while of the hosts that show emission lines, only 3 have a clear AGN signature in a BPT diagram. This provides supporting evidence that the majority of these events are indeed due to the tidal disruptions of stars, and not due to accretion disk instabilities or alternative scenarios related to AGN activity. We further analysed the host galaxy properties, and conclude the following:

\begin{itemize}
\item We identify two new members of the quiescent moderately Balmer strong class of galaxies: iPTF--16axa (optical TDE) and LEDA 095953 (soft X-ray TDE). For a third candidate with strong H$\delta_A$ absorption, DES14C1kia, the H$\alpha$ line is redshifted into an atmospheric band and we cannot verify the lack of H$\alpha$ emission (although no other lines expected in star-forming galaxies are detected).\\

\item Three hard X-ray TDE candidates (out of a total sample size of 6 for this class) occurred in dwarf galaxies with M$_{\rm g}$\,$\sim$\,--16 and M$_{*}$\,$\sim$\,10$^{8.5-9}$, an order of magnitude fainter and less massive than the least massive optical TDE host. It is still unclear whether these events are truly TDEs. Two soft X-ray host galaxies fall in the gap between these two groups. Although the black hole masses are uncertain, this further indicates that the rate of TDEs is dominated by low mass BHs and host galaxies, as predicted by theory. If the hard X-ray events are indeed TDEs, we conclude that a non-zero fraction of dwarf galaxies at those masses host massive black holes. For these three galaxies, the black hole mass derived from the velocity dispersion is $\sim$\,10$^{4.5-5}$\,M$_{\odot}$; if instead we use the galaxy stellar mass, we find M$_{\rm BH}$\,$\sim$\,10$^{5.5}$\,M$_{\odot}$.\\ 

\item There is no robust statistical evidence that the TDE host SMBH masses, stellar masses, or absolute magnitudes are drawn from different parent distributions when one compares our UV/optical and soft X-ray selected subsamples. For the hard X-ray sample, on the other hand, we can reject the hypothesis that its host properties are drawn from the same parent distribution as the optical and/or soft X-ray samples at the $\sim$\,2\,$\sigma$ level. The fact that this is the case for all three host properties we examine suggests that this difference is likely real, although a larger sample should be used to provide conclusive evidence. \\

\item The Eddington ratio of the observed TDE X-ray emission is typically of order 0.01 (with a large range 10$^{-4}$\,--\,1), whereas for the optical emission it is typically of order 1. There is no correlation between L$_{\rm Edd,X}$ and M$_{\rm BH}$, which might have been expected were AGN outbursts or unobscured accretion of rapidly circularizing TDE debris the power source. Instead, the Eddington ratio of the X-ray emission is similar to that of late-time UV emission \citep{vanvelzen2018b}. This implies that the accretion rate at the time of the X-ray observations is similar to the accretion rate at the time of the UV measurements (although potential obscuration and/or a bolometric correction are not yet taken into account).\\

\item Estimates of the emission region of the X-ray radiation leads us to conclude that it is (as expected) close to the black hole, and significantly closer-in than both the early and late-time UV/optical emission regions. The X-ray spectra suggest accretion disk emission, and a compact accretion disk model is certainly consistent with the observed Eddington ratios (the same is true for late-time UV emission). The size of the X-ray emitting region is more puzzling, however. While our simple fits for emitting area are subject to several uncertainties, if taken at face value, they indicate that many soft X-ray TDEs suffer from a high level of obscuration. Future work that more carefully examines uncertainties in the original X-ray spectrum could therefore usefully test the presence of reprocessing layers.\\

Many of our conclusions have been limited by the small size of the current TDE candidate sample. In the near future, this sample will expand by two orders of magnitude, as eROSITA and LSST begin discovering thousands of new TDE candidates. This near-future sample will contain large amounts of information on the SMBH mass function, and in particular its low-end occupation fraction. However, such a large number of host galaxies may frustrate attempts to generalise the relatively expensive spectroscopic work of this study, highlighting the need for improved TDE light curve models that can infer SMBH masses through flare photometry alone.

\end{itemize}

%%%%%%%%%%%%%%%%%%%%%%%%%%%%%%%%%%%%%%%%%%%%%%%%%%%%%%%%%%%%%%%%%%%%%%%%%%%%%%%%%%%%%%%%%%%%%%%%%%%%%%%%%%%%%%%%%%%%%%%%%%%%%%%%%%%%%%%%%%%%%%%%%%%%%%%%%%%%%%%%%%%%%%%%%%%%%%%%%%%%%%%%%%%%%%%%%%%%%%%%%%%%%%%%%%%%%%%%%%
\section*{Acknowledgements}
We thank the referee for thoughtful and constructive comments. TW thanks M. Auger, P. Hewett and R. Saxton for useful discussions and suggestions, and D. Lena, T. van Grunsven and E. Breedt for performing part of the observations. TW is funded in part by European Research Council grant 320360 and by European Commission grant 730980. 
SG is supported in part by NSF CAREER grant 1454816 and NASA Keck Grant 1568615. PGJ, FO and ZKR acknowledge funding from the European Research Council under ERC Consolidator Grant agreement no 647208. NCS acknowledges funding from NASA ATP grant NNX17AK43G and NASA Einstein Postdoctoral Fellowship Award Number PF5-160145. D.M.S. acknowledges support from the ERC under the European Union's Horizon 2020 research and innovation programme (grant agreement No. 715051; Spiders). JC acknowledges support by the Spanish Ministry of Economy, Industry and Competitiveness (MINECO) under grantAYA2017-83216-P. Some of the data presented herein were obtained at the W.M. Keck Observatory, which is operated as a scientific partnership among the California Institute of Technology, the University of California and the National Aeronautics and Space Administration. The Observatory was made possible by the generous financial support of the W.M. Keck Foundation. We wish to recognise and acknowledge the very significant cultural role and reverence that the summit of Mauna Kea has always had within the indigenous Hawaiian community. We are most fortunate to have the opportunity to conduct observations from this mountain. The ISIS spectroscopy was obtained as part of programmes W15BN10, W16AN7, W16BN4, W16BN7, W17AN6, W17BN6 and W18AN4. The WHT is operated on the island of La Palma by the Isaac Newton Group of Telescopes in the Spanish Observatorio del Roque de los Muchachos of the Instituto de Astrofisica de Canarias. Based on observations made with the Gran Telescopio Canarias (GTC, proposal id GTC71-17A), installed in the Spanish Observatorio del Roque de los Muchachos of the Instituto de Astrofísica de Canarias, in the island of La Palma.

\bibliographystyle{mnras.bst}
\bibliography{bibliography_tdehosts}

%%%%%%%%%%%%%%%%%%%%%%%%%%%%%%%%%%%%%%%%%%%%%%%%%%%%%%%%%%%%%%%%%%%%%%%%%%%%%%%%%%%%%%%%%%%%%%%%%%%%%%%%%%%%%%%%%%%%%%%%%%%%%%%%%%%%%%%%%%%%%%%%%%%%%%%%%%%%%%%%%%%%%%%%%%%%%%%%%%%%%%%%%%%%%%%%%%%%%%%%%%%%%%%%%%%%%%%%%%

\appendix
\setcounter{table}{0}
\renewcommand{\thetable}{A\arabic{table}}
\section{Additional tables and observing log}
In this Appendix we provide two additional tables. Table \ref{tab:xraymeasurements} provides the details of the X-ray spectral fitting, and Table \ref{tab:obslog} provides the observing log for all the spectra used in this work as well as those of \citet{Wevers2017}.
\label{sec:appendix}
\begin{table*}
\caption{In this table we give the relevant details of the blackbody model fitting of the X-ray spectra. Obs ID gives the observation identifier, MJD is the modified julian date, kT is the blackbody temperature, bbnorm is the normalisation constant of the blackbody spectrum, the radius is the physical radius of the emission region, r$\chi^2$ is the reduced $\chi^2$ statistic of the fit and d.o.f gives the number of degrees of freedom.}
  \begin{tabular}{lcccccccc}
  \hline
Name & Satellite & Obs ID & MJD & kT (eV) & bbnorm& Radius (cm) & r$\chi^2$ & d.o.f  \\
  \hline\vspace{0.5mm}
ASASSN--14li & XMM & x0722480201 & 56999.5 & 51.2$^{+0.1}_{-0.1}$ & 3.21$^{+0.42}_{-0.36}$$\times$10$^6$ & 1.6$^{+0.1}_{-0.1}$$\times$10$^{12}$ & 3.1 & 100  \\\vspace{0.5mm}
SDSS J0159 & XMM &  x0101640201 & 51754.9 & 423$^{+52}_{-44}$ & 1.5$^{+0.8}_{-0.5}$& 2.0$^{+0.5}_{-0.4}$$\times$10$^{10}$& 3.1 & 25\\\vspace{0.5mm}
2MASX J0249 & XMM & x0411980401 & 53930.5 & 77.5$^{+0.3}_{-0.3}$ & 5.8$^{+2.5}_{-1.7}$$\times$10$^4$& 2.0$^{+0.4}_{-0.3}$$\times$10$^{11}$& 1.7 & 54 \\\vspace{0.5mm}
SDSS J1201 & XMM & x0555060301 & 55369.4 & 178$^{+6}_{-6}$ & 50$^{+9}_{-8}$& 5.0$^{+0.5}_{-0.4}$$\times$10$^{10}$& 1.4 & 127 \\\vspace{0.5mm}
RBS 1032 & ROSAT & rp201237n00 & 49000 & 26.5$^{+0.7}_{-0.5}$ &2.9$^{+15}_{-3.1}$$\times$10$^6$ &2$^{+5}_{-1}$$\times$10$^{12}$ & 1.4 & 15 \\\vspace{0.5mm}
ASASSN--15oi & Swift & sw00033999001--sw00033999012 & 57275.1 & 36$^{+10}_{-10}$ &1.6$^{+27}_{-2.4}$$\times$10$^5$ & 1.1$^{+0.5}_{-0.5}$$\times$10$^{11}$& 0.7 & 20 \\\hline
  \end{tabular}
    \label{tab:xraymeasurements}
  \end{table*}
  
 \begin{table*}
 \caption{In this table we provide the observational setups, observing dates and exposure times of the new optical long-slit spectra presented in this work. In addition, we also provide this information for the spectra presented in \citet{Wevers2017}.}
 \begin{tabular}{lcccc}
 \hline 
Name &Telescope& Instrument/Grating & Observing date & Exposure time (s) \\\hline
2MASX J02491731-0412521& WHT & ISIS/R600 & 2017--08--27 & 6x600\\
2MASX J02491731-0412521& Keck & ESI & 2017--11--17 & 300 \\ 
3XMM J150052.07+015453.8 & GTC & Osiris/R2500V & 2017--05--20 & 2700s \\
3XMM J152130.72+074916 & GTC & Osiris/R2500V & 2017--05--21 & 2x2700s \\
ASASSN--14li & WHT & ISIS/R600 & 2016--07--03& 2x1200 \\
ASASSN--14li & Keck & ESI & 2017--02--24& 360 \\
ASASSN--14ae & WHT & ISIS/R600 & 2016--07--02& 2x1200 \\
ASASSN--14ae & Keck & ESI & 2017--02--24& 700 \\
ASASSN--15oi & WHT & ISIS/R600 & 2017--08--20& 4x900\\
ASASSN--15oi & WHT & ISIS/R600 & 2017--09--29& 4x2700\\
DES14--C1kia & Keck & ESI & 2017--11--17 & 1000 \\
GALEX D19 & Keck & ESI & 2017--11--17 & 2x1800\\
GALEX D23H1& WHT & ISIS/R600 & 2015--08--10&5x2700 \\
GALEX D23H1& WHT & ISIS/R600&2015--09--14&3x2700\\
GALEX D23H1& Keck & ESI & 2017--11-17 & 1500\\
iPTF--15af& Keck & ESI & 2017--02--24& 900 \\
iPTF--16axa& Keck & ESI & 2017--02--24& 1800 \\
iPTF--16fnl & VLT&XSHOOTER&2016--11--01&590\\
2MASX J14460522+6857311 & WHT & ISIS/R600 & 2018--07--08 & 2700 + 2050\\
LEDA 095953 & WHT& ISIS/R300 & 2016--02--17 & 2x1800\\
NGC 5905 & WHT & ISIS/R600 & 2017--03--09 & 6x600\\
NGC 6021 & WHT&ISIS/R300&2016--02--17 & 1200\\
PGC 015259 & WHT & ISIS/R300 & 2016--02--16 & 1200\\
PGC 1127938 & Keck & ESI & 2017--11--17 & 1000\\
PGC 133344 & WHT & ISIS/R600 & 2017--08--28 & 2x2400\\
PGC 170392 & WHT & ISIS/R600 & 2016--07--03 & 2x1200\\
PGC 170392 & WHT & ISIS/R600 & 2017--08--29 & 6x600\\
PS1--10jh& Keck & ESI & 2017--02--24& 3600 \\
PS1--10jh& GTC & Osiris/R2500V & 2017--05--19 & 2700s \\
PS1--10jh& GTC & Osiris/R2500V & 2017--05--20 & 2x2700s \\
PS1--11af& Keck & ESI & 2017--11--17 & 2x1800\\
PTF--09axc & WHT&ISIS/R600&2015--09--13&2700\\
PTF--09djl& WHT&ISIS/R600&2015--08--09&3x2700\\
PTF--09djl&WHT&ISIS/R600&2015--09--13&2x2700\\
PTF--09djl& Keck & ESI & 2017--02--24& 700 \\
PTF--09ge& Keck & ESI & 2017--02--24& 600 \\
PTF--09ge& WHT & ISIS/R600&2015--09--13&2x1200\\
RX J1242.6​-1119A & WHT & ISIS/R600&2018--07--08 & 1800 \\
RX J1624.9​+7554 & WHT&ISIS/R600&2015--08--10 & 600\\
RX J1624.9​+7554 & WHT&ISIS/R600&2015--09--13 & 600\\
SDSS J015957.64​+003310.4& Keck & ESI & 2017--11--17 & 1000\\
SDSS J120136.02​+300305.5& Keck & ESI & 2017--11--17 & 600\\
SDSS TDE1 & WHT & ISIS/R600 & 2015--09--14 & 4x2700\\
SDSS TDE2 & Keck & ESI & 2017--11--17 & 2000\\
UGC 1791 & Keck & ESI & 2017--11--17 & 1000\\\hline
 \end{tabular}
   \label{tab:obslog}
 \end{table*}

%%%%%%%%%%%%%%%%%%%%%%%%%%%%%%%%%%%%%%%%%%%%%%%%%%%%%%%%%%%%%%%%%%%%%%%%%%%%%%%%%%%%%%%%%%%%%%%%%%%%%%%%%%%%%%%%%%%%%%%%%%%%%%%%%%%%%%%%%%%%%%%%%%%%%%%%%%%%%%%%%%%%%%%%%%%%%%%%%%%%%%%%%%%%%%%%%%%%%%%%%%%%%%%%%%%%%%%%%%

\label{lastpage}
\end{document}